\documentclass[aps,pra,10pt,floatfix,twocolumn,showpacs,floatfix,superscriptaddress,longbibliography]{revtex4-2}

\usepackage{CJK} 

\usepackage{amsthm,amsmath,amssymb,amsfonts,bbm} 
\usepackage{mathrsfs} 
\usepackage{lipsum} 

\usepackage{graphicx} 
\usepackage{subfigure}

\usepackage{xcolor}
\usepackage[colorlinks=true,citecolor=blue,urlcolor=blue,linkcolor=red,hyperindex]{hyperref} 

\usepackage{tikz,tikz-3dplot,tikz-dependency}
\usetikzlibrary{quotes,angles,arrows,arrows.meta,positioning,calc,snakes,fadings,shadings,intersections,decorations.text}
\usepackage{overpic} 

\begin{document}

\title{Parity flipping mediated by a quantum dot in Majorana Josephson junctions}

\author{Shanbo Chow}
\affiliation{School of Physics, State Key Laboratory of Optoelectronic Materials and Technologies, Sun Yat-sen University, Guangzhou 510275, China}

\author{Zhi Wang}
\affiliation{School of Physics, State Key Laboratory of Optoelectronic Materials and Technologies, Sun Yat-sen University, Guangzhou 510275, China}

\author{Dao-Xin Yao}
\email{yaodaox@mail.sysu.edu.cn}
\affiliation{School of Physics, State Key Laboratory of Optoelectronic Materials and Technologies, Sun Yat-sen University, Guangzhou 510275, China}
\affiliation{International Quantum Academy, Shenzhen, 518048, Guangdong, China}

\begin{abstract}
The detection of the Majorana bound states (MBSs) is a central issue in the current investigation of the topological superconductors, and the topological Josephson junction is an important system for resolving this issue. In this work, we introduce an external quantum dot (QD) to Majorana Josephson junctions (MJJs), and study the parity flipping of the junction induced by the coupling between the QD and the MBSs. We demonstrate Landau-Zener (LZ) transitions between opposite Majorana parity states when the energy level of the QD is modulated. The resulted parity flipping processes exhibit voltage signals across the junction. In the presence of a periodic modulation on the QD level, we show Landau-Zener-St\"{u}ckelberg (LZS) interference on the parity states. We demonstrate distinctive interference patterns at distinct driving frequencies. These results can be used as signals for detecting the existence of the MBSs.
\end{abstract}

\date{\today}

\maketitle
\section{Introduction}\label{Introduction}
Recently, MBSs in topologically superconducting systems have attracted much attention due to the non-Abelian statistics, braiding operations and possible applications in topological quantum computation \cite{albrecht2016exponential, doi:10.7566/JPSJ.85.072001, Zhu_2016, Wang_2020, KITAEV20032, RevModPhys.80.1083, Field_2018, RevModPhys.73.357, PhysRevLett.103.020401, PhysRevLett.128.016402}. Theoretical proposals for MBSs are mostly based on heterostructures, where topological nontrivial states appear with the superconducting proximity effect and a delicate balance of system parameters \cite{PhysRevLett.100.096407, RevModPhys.83.1057}. With the rapid progresses in material growth technologies, detection of the MBSs becomes an urgent issue. While experimental signatures for MBSs have been reported in a number of experiments \cite{PhysRevB.61.10267, PhysRevLett.105.077001, doi:10.1126/science.1222360, PhysRevLett.114.017001, Alicea_2012, PhysRevLett.115.177001, PhysRevLett.100.096407, doi:10.1126/science.1259327}, evidence for MBSs is still required for further pursuit of topological quantum computation.

Among the various systems for MBSs, the topological Josephson junction is particularly interesting. A typical one-dimensional topological junction hosts two MBSs at the two sides of the junction. These two MBSs couple with each other, defining a two-level system where the two states represent the even and odd fermionic parities of the junction \cite{KITAEV20032, PhysRevB.73.014505, 10.21468/SciPostPhys.12.5.161, zhou2022fusion}. The even/odd parity defines the two distinct ground state of the junction and is responsible for the $4\pi$-periodic Josephson current phase relation. The quantum dynamics of the parity state induced by the injection current has been shown to be useful for understanding the transport features of the topological junctions.

For topological superconductors, it has been shown that an external QD, such as epitaxial region of semiconducting InAs wire \cite{PhysRevB.96.195430}, can flip the parity of the topological ground states. QDs are zero dimensional systems which can be described by a single fermionic level. When the energy of this QD level is raised from deep below zero to well above zero, one electron would hop from the QD to the nearby MBSs and the parity state is flipped. One would expect interesting features if the parity state of a topological Josephson junction is flipped with a QD. In particular, a dynamical manipulation of the QD level would induce transport response in the Josephson junction.

In this work, we study the voltage evolution induced by parity flipping through LZ transition in a MJJ as sketched in Fig. \ref{fig:setup_QD}. Four MBSs appear in the junction, with two of them located at the ends of the wire and the other two located near the tunneling barrier in the middle of the wire. The two MBSs near the tunneling barrier define a local Majorana parity state \cite{Kitaev_2001, PhysRevLett.86.268}, and the QD is only coupled to the second MBS from left to right, thus the coupling between them generates a $4 \times 4$ Hamiltonian which determines quantum dynamics of the parity state through a Schr\"{o}dinger equation. The Josephson Hamiltonian is critically influenced by the superconducting phase difference across the junction, which is well described by the quantum resistively and capacitively shunted junction (RCSJ) model \cite{PhysRevB.98.134515, PhysRevB.101.180504, Xu_2017, PhysRevB.102.140501, PhysRevB.103.094518}. In this model, the classical dynamics of the phase difference is determined by the quantum average of local Majorana parity. This combination of classical dynamics of the phase difference and quantum dynamics of the Majorana parity state yields a junction dynamics distinct from conventional RCSJ model without Majorana channel. We calculate the time evolution of the Josephson phase with a raising QD level energy, and find that the quantum dynamics of the Majorana parity state yields a 4$\pi$-period voltage for MJJs. Then we find a $4\pi$-periodic hopping behavior of the QD occupation probability curve after parity flipping in the underdamped regime. These behaviors can be used as possible signatures for the existence of MBSs since it is unique to MJJs. Meanwhile, we consider the influence of the decoherence to the voltage curve. Next, we study the LZS interferences at distinct driving frequencies in our set-up. Then we find it can realize the switch between voltage and voltage-free states by tuning the driving frequency. Finally, our result of the Rabi oscillation is similar to some damped Rabi oscillations observed in experiments \cite{doi:10.1063/1.5078628, Schon2020, Shevchenko_2008, PhysRevLett.95.067001, PhysRevB.78.104510}, which can be referred in quantum circuit design.

This work is organized as follows. The model for MJJs is introduced in Section \ref{Section:Set-up Model}. The voltage induced by parity flipping for zero temperature regime and the impact of decoherence on it are shown in Section \ref{Section:Landau-Zener transition}. The pattern of LZS interference is studied in Section \ref{Section:LZS interference}, and the result of the damped Rabi oscillation is included in Section \ref{Section:Rabi oscillation}. Finally, we give conclusions in Section \ref{Section:Conclusions}.

\section{Set-up Model}\label{Section:Set-up Model}
\subsection{Four-Level System}
\begin{figure}[h!]
	\centering
	\begin{overpic}[scale=0.2]{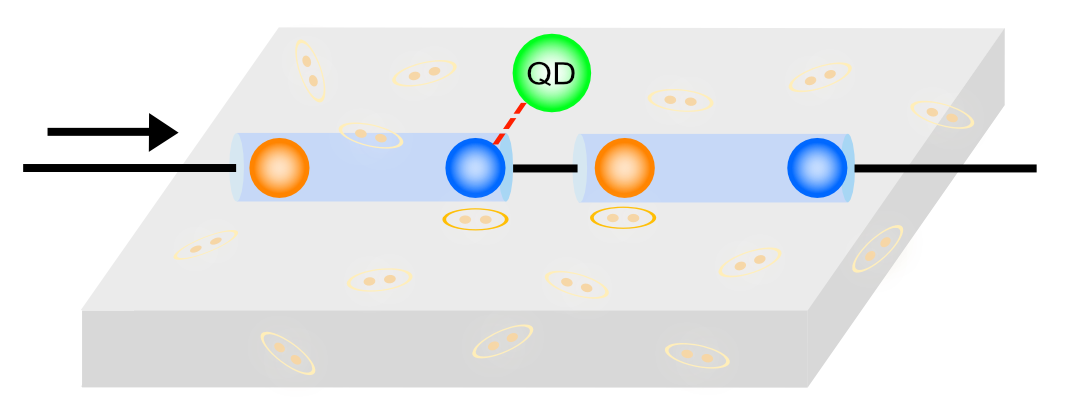}
		\put(8,28){\small$I$}
		\put(24,28){\small$\gamma_1^{}$}
		\put(42,28){\small$\gamma_2^{}$}
		\put(56,28){\small$\gamma_3^{}$}
		\put(74,28){\small$\gamma_4^{}$}
		\put(20,5){\small$s$-wave superconductor}
	\end{overpic}
	\caption{Schematic of the MJJ with a QD.}
	\label{fig:setup_QD}
\end{figure}
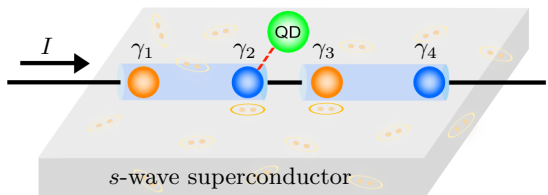

The MJJ sketched in Fig. \ref{fig:setup_QD} consists of a spin-orbit coupling nanowire placed over a bulk conventional $s$-wave superconductor where the supercurrent is contributed by tunneling in the form of both single electron and Cooper pair. The nanowire is adjusted to the topological states with the appropriate chemical potential and Zeeman field. A voltage gate is applied in the middle of the wire to create a tunneling barrier. There are four localized MBSs within the wire. Two of them, $\gamma_1^{}$ and $\gamma_4^{}$, are locating at the two ends of the wire, while the other two, $\gamma_2^{}$ and $\gamma_3^{}$, reside at the two sides of the tunneling barrier, respectively. Then we can define two fermion operators $f_1^{} = (\gamma_2^{} + i\gamma_3^{})/2$, $f_2^{} = (\gamma_4^{} + i\gamma_1^{})/2$ \cite{majorana1937teoria}. These two MBSs near the barrier provide a novel channel for electron tunneling, which introduces an extra $2 \times 2$ Josephson Hamiltonian in addition to the conventional Josephson energy from the finite energy quasiparticle channels. For simplicity, we put a QD coupling to only $\gamma_2^{}$. Including all these aspects, the total Hamiltonian of the system can be separated to four parts:
\begin{equation}\label{eq:Hamiltonian_total}
	H
	= H_{\mathrm{J}}^{}
	+ H_\delta^{}
	+ H_\mathrm{QD}^{}
	+ H_\mathrm{T}^{}.
\end{equation}
The Josephson Hamiltonian caused by the tunneling of Cooper pair and single electron is
\begin{align}\label{eq:Hamiltonian_J}
	H_{\mathrm{J}}^{}
	&= H_{\mathrm{C}}^{} + H_{\mathrm{M}}^{} \notag\\
	&= -E_\mathrm{J}^{} \cos{\theta} + i E_\mathrm{M}^{} \gamma_2^{}\gamma_3^{} \cos({\theta}/{2})
\end{align}
where $E_\mathrm{J}^{}$ and $E_\mathrm{M}^{}$ account for the amplitude of $2\pi$- and $4\pi$-periodic Josephson energy respectively, and $\theta$ is the gauge invariant phase difference across the junction. The first term in the Josephson Hamiltonian derives from the finite energy quasiparticle channels, which resembles the traditional $2\pi$-period Josephson effect. Note that this term could be omitted because it has no contribution to the parity flipping since it can be expanded to an identity matrix up to scalar multiplication. The second term comes from the tunneling between MBSs which leads to a novel $4\pi$-periodic Josephson energy known as the fractional Josephson effect \cite{kwon2004fractional, rokhinson2012fractional}.

The hybridization Hamiltonian can be written as \cite{Kitaev_2001}
\begin{equation}\label{eq:Hamiltonian_hybridization}
	H_\delta^{} = i\delta_\mathrm{L}^{} \gamma_1^{}\gamma_2^{} + i\delta_\mathrm{R}^{} \gamma_3^{}\gamma_4^{}
\end{equation}
where $\delta_{\mathrm{L}(\mathrm{R})}^{}$ are the small couplings between the MBSs in the left (right) side of the tunneling barrier.

The third Hamiltonian is the QD Hamiltonian
\begin{equation}\label{eq:Hamiltonian_QD}
	H_{\mathrm{QD}}^{} = \varepsilon d^{\dagger}d,
\end{equation}
here $d^{\dagger}$, $d$ are the creation and annihilation operators of QD, and $\varepsilon$ is the energy of the QD.

The QD can couple with one or some of the four Majorana modes. Here, we choose the simplest case, i.e., the QD only couples with $\gamma_2^{}$. Hence the standard tunneling Hamiltonian through $\gamma_2^{}$ to QD is
\begin{align}\label{eq:Hamiltonian_tunneling}
H_\mathrm{T}^{}
	&= T_\mathrm{L}^{} \gamma_2^{} d - T_\mathrm{L}^* d^{\dagger} \gamma_2^{} \notag\\
	&= (f_1^{}+f_1^{\dagger}) (T_\mathrm{L}^{} d - T_\mathrm{L}^* d^{\dagger})
\end{align}
where $T$ and $T^*$ are the tunneling quantities.

After including a QD, we can define a four-state system with four odd parity states $f_2^{\dagger}|00\rangle = |1\rangle$, $f_1^{\dagger}|00\rangle = |2\rangle$, $d^{\dagger}|00\rangle = |3\rangle$, $d^{\dagger}f_2^{\dagger}f_1^{\dagger}|00\rangle = |4\rangle$ due to the conservation of the total parity. Typically, $\delta_{\mathrm{L}(\mathrm{R})}^{} \ll E_{\mathrm{M}}^{}$ as these couplings between MBSs decay exponentially when the length of the wire increases. The increase of the QD energy would break the local parity conservation of the Josephson Hamiltonian, and drives quantum evolution of the parity state through a Schr\"{o}dinger equation $i \frac{{\rm d}}{{\rm d}t} |\Psi(t)\rangle = H |\Psi(t)\rangle$, where $H = H_{\mathrm{M}}^{} + H_\delta^{} + H_\mathrm{QD}^{} + H_\mathrm{T}^{}$ and $|\Psi(t)\rangle = \sum_n^4 c_n^{}(t) |n\rangle = \sum_n^4 e^{-iE_n^{}t/\hbar} c_n^{}(0) |n\rangle$. Hence, the total Hamiltonian $H$ can be expanded to a $4 \times 4$ matrix
\begin{equation}\label{eq:unreduced total Hamiltonian1}
{\footnotesize
	\mathcal{H}
	=
	\begin{pmatrix}
		E_\mathrm{M}^{}\cos{\frac{\theta}{2}} & \delta_1^{} & 0 & T^* \\
		\delta_1^{} & -E_\mathrm{M}^{}\cos{\frac{\theta}{2}} & -T^* & 0 \\
		0 & -T & E_\mathrm{M}^{}\cos{\frac{\theta}{2}}+\varepsilon(t) & \delta_2^{} \\
		T & 0 & \delta_2^{} & -E_\mathrm{M}^{}\cos{\frac{\theta}{2}+\varepsilon(t)} \\
	\end{pmatrix}
}
\end{equation}
where $E_\mathrm{M}^{} = 20 \mathrm{~\mu eV}$, $\delta_\mathrm{L}^{} = 0.0015 E_\mathrm{M}^{}$, $\delta_\mathrm{R}^{} = 0.0005 E_\mathrm{M}^{}$, and $T = T^* = 0.01 E_\mathrm{M}^{}$. For simplicity, we set $\delta_1^{} = -(\delta_\mathrm{L}^{} + \delta_\mathrm{R}^{})$, $\delta_2^{} = -(\delta_\mathrm{L}^{} - \delta_\mathrm{R}^{})$.

\subsection{Total Supercurrent}
The MJJ has an internal degree of freedom, i.e., the parity state of the MBSs. This parity state is a unique feature of MJJs, which distinguishes it from conventional Josephson junctions. More importantly, quantum dynamics of this parity state is controlled by the phase difference $\theta$, indicating a correlation between the parity degree of freedom and the phase degree of freedom.

Now we start to explore the influence of the parity state on the phase difference. Dynamics of the phase difference can be described by a RCSJ model. Within the RCSJ model, the injection current $I_{\mathrm{in}}^{}$ goes through four channels --- the capacitance, resistance, Cooper pair, and single electron channels. Therefore the current conservation gives
\begin{equation}\label{eq:unreduced current}
	I_{\mathrm{in}}^{} = \dfrac{\hbar C}{2e}\ddot{\theta} + \dfrac{\hbar}{2eR}\dot{\theta} + I_1^{} \sin{\theta} + I_2^{} S_z^{}\sin(\theta/2)
\end{equation}
where the Josephson current and the Majorana current are
\begin{equation*}
	I_\mathrm{J}^{} = I_1^{} \sin\theta, \qquad I_\mathrm{M}^{} = I_2^{} S_z^{} \sin(\theta/2),
\end{equation*}
and $I_1^{} = 6.3 \mathrm{~nA}$, $I_2^{} = 1.6 \mathrm{~nA}$, $I_{\mathrm{in}}^{} = 7 \mathrm{~nA}$. Note that $I_{\mathrm{in}}^{}$ can not be chosen arbitrarily. First, the resistance and capacitance is adjusted properly to make the MJJ circuit in the overdamped regime with the quality factor $Q = R\sqrt{2eCI_\mathrm{uc}^{}/\hbar} < 1$. Then we can find a switching current $I_\mathrm{sw}^{}$ and a smaller recapturing current $I_\mathrm{re}^{}$ via plotting the hysteresis curve through an upward and downward scanning current. Finally, we inject the constant current bias $I_\mathrm{in}^{}$, which is required to be above the lower critical current $I_\mathrm{lc}^{} = I_\mathrm{re}^{}$ and slightly below the upper critical current $I_\mathrm{uc}^{} = I_\mathrm{sw}^{} = \max(I_\mathrm{J}^{} + I_\mathrm{M}^{}) \approx 7.46 \mathrm{~nA}$, to our set-up \cite{PhysRevB.98.134515}. The other two parameters, the capacitance $C \approx 0.659 \mathrm{~nF}$ and the resistance $R = 50 ~\Omega$, also are realistic values close to the estimations in the experiments \cite{PhysRevX.7.021011, PhysRevB.101.180504}. Note that the system now is in the underdamped regime with $Q>1$ instead of the early overdamped regime for searching the upper and lower critical currents since the hysteresis can occur in overdamped MJJs \cite{PhysRevB.98.134515, PhysRevB.101.180504}.

The Majorana operators mentioned above can define a parity operator $\hat{s}_z^{} = i\gamma_2^{} \gamma_3^{}$. If this local Majorana parity is conserved, the Hamiltonian could be simplified to a $4\pi$-periodic Josephson energy, on which a phase difference derivative (proportional to the induced voltage) could appear a $4\pi$-period. There is an intrinsic parity flipping process existing at zero temperature, which can be described by the relative amplitude $S_z^{}$,
\begin{align}\label{eq:relative amplitude}
	S_z^{}(t) &= \langle\Psi(t)| \hat{s}_z^{} |\Psi(t)\rangle
	= \langle\Psi(t)| i\gamma_2^{} \gamma_3^{} |\Psi(t)\rangle \notag\\
	&= - |c_1^{}(t)|^2 + |c_2^{}(t)|^2 - |c_3^{}(t)|^2 + |c_4^{}(t)|^2.
\end{align}

\section{Landau-Zener transition}\label{Section:Landau-Zener transition}
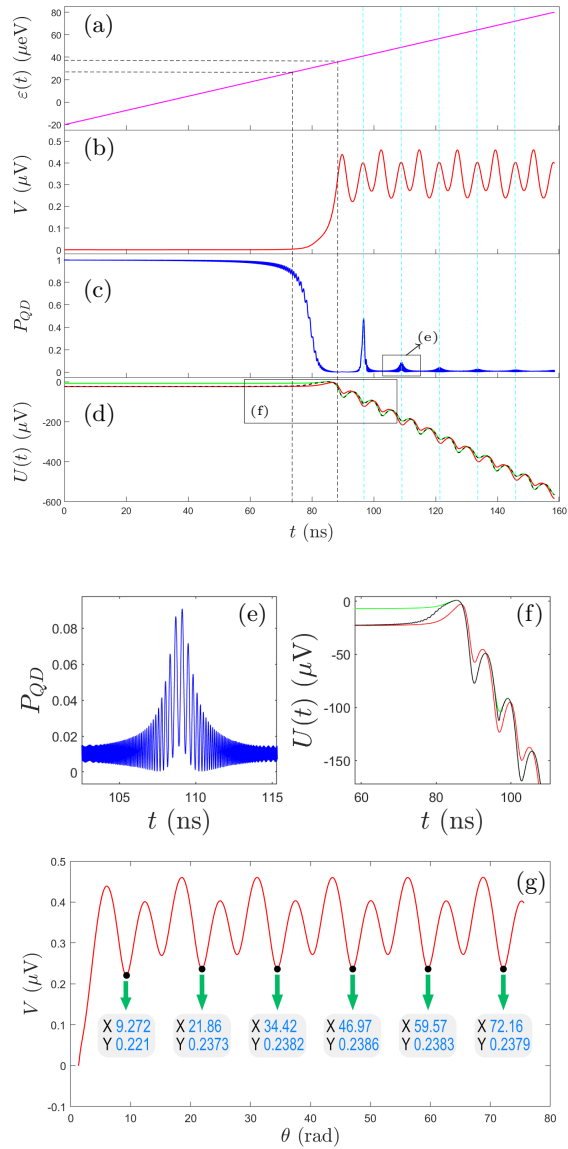
\begin{figure}[t!]
\begin{center}
	\begin{overpic}[clip = true, width = 0.95 \columnwidth]{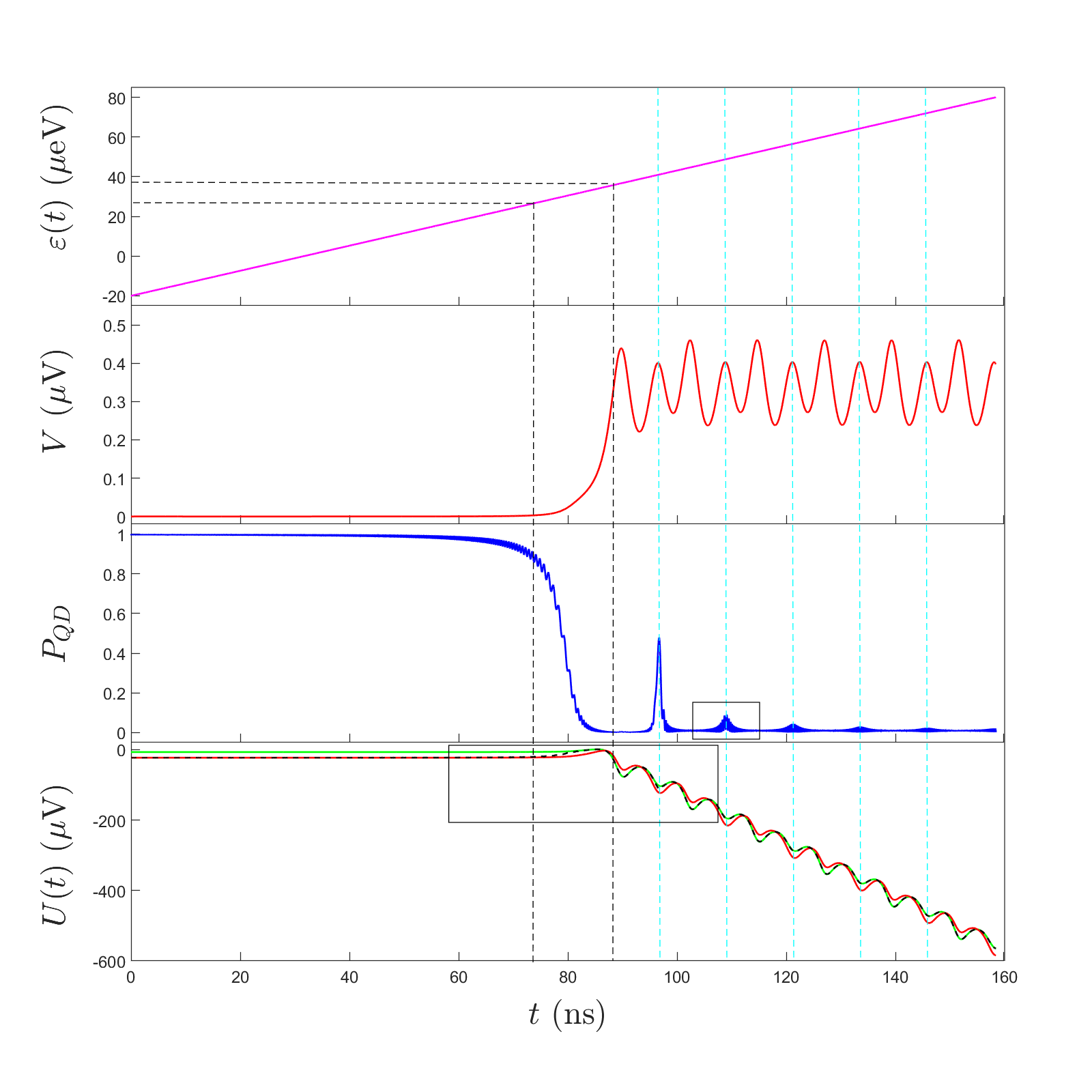}
		\put(15,88){\small (a)}
		\put(15,68){\small (b)}
		\put(15,45){\small (c)}
		\put(15,25){\small (d)}
		\put(67,36){\tiny $\nearrow$}
		\put(69,38){\tiny (e)}
		\put(42,26){\tiny (f)}
	\end{overpic}
	\begin{overpic}[clip = true, width = 0.95 \columnwidth]{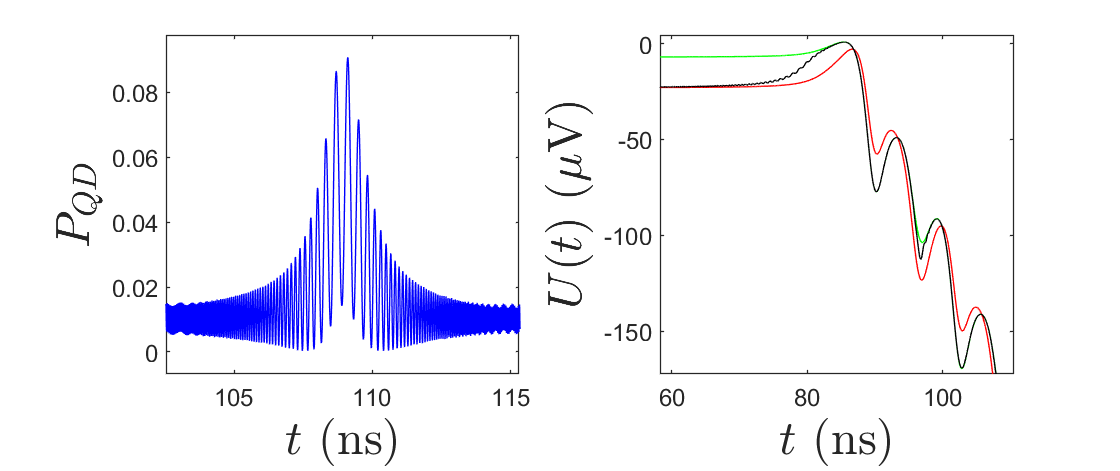}
		\put(40,35){\small (e)}
		\put(85,35){\small (f)}
	\end{overpic}
	\begin{overpic}[clip = true, width = 0.95 \columnwidth]{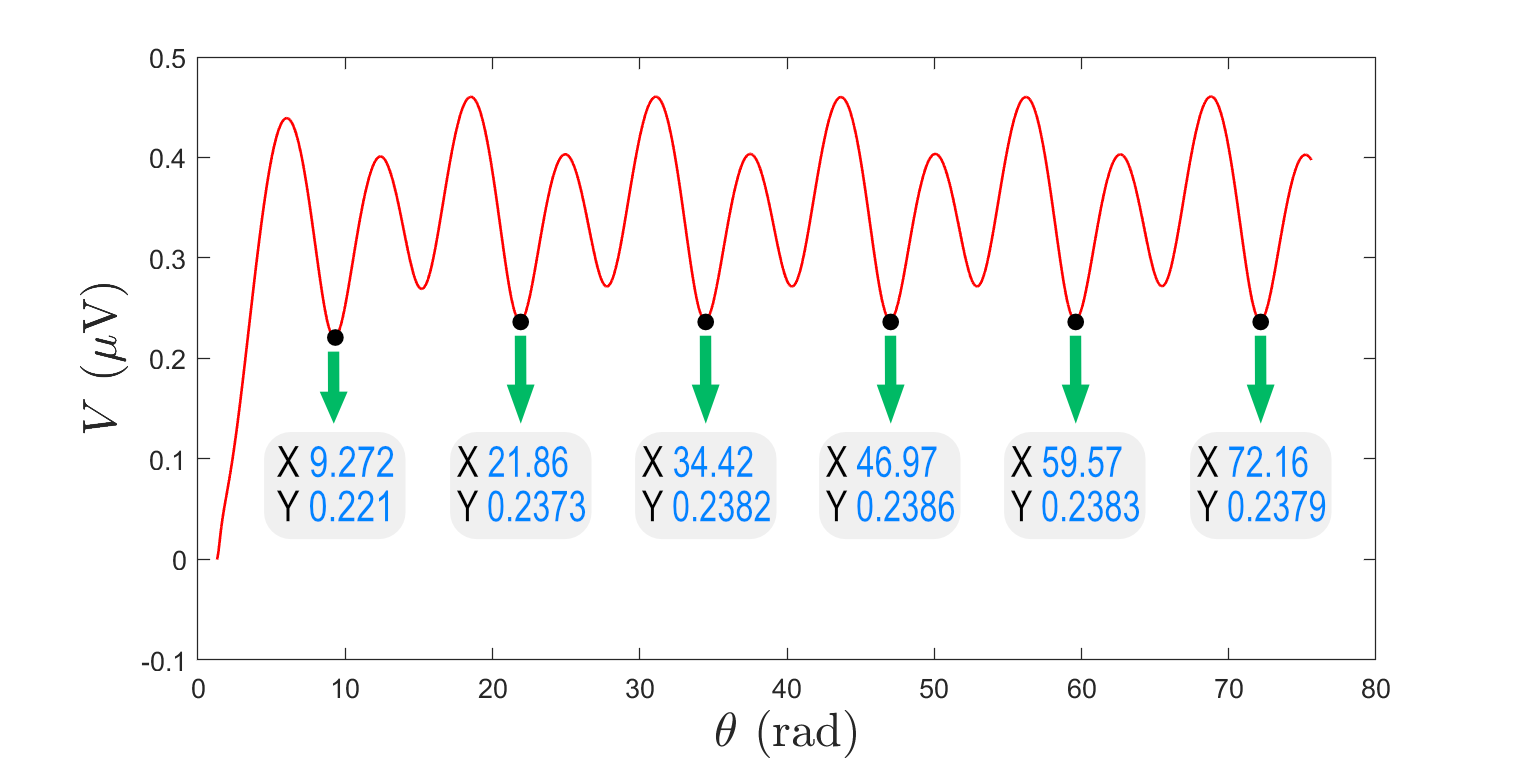}
		\put(85,42){\small (g)}
	\end{overpic}
	\caption{The case of fast evolution. The tunneling coefficient $T = 0.1 E_{\mathrm{M}}^{}$ and the $\varepsilon(t)$ ranges from $-E_{\mathrm{M}}^{}$ to $4 E_{\mathrm{M}}^{}$. (a) The energy of the QD. (b)(g) The time-voltage and phase-voltage curves. They show a $4\pi$-periodic oscillation from MBSs. (c) The occupation probability of the QD and (d) the effective tilted washboard potential $U(\theta) = -E_\mathrm{J}^{} \cos{\theta} - E_\mathrm{M} S_z^{} \cos{\tfrac{\theta}{2}} - (\hbar I/2e)\theta$. In this process, the parity flips from $1$ to $-1$ which corresponds to the change from the lower potential (red line) to the higher one (green line). It also shows a $4\pi$-period related to MBSs. (e) The fast wiggle of the QD occupation probability has small amplitudes which arises from the Larmor precession. (f) The close-up of the effective tilted washboard potential, from the beginning of the parity flipping process to the second peak, explicitly presents the QD occupation approach.}
	\label{fig:I-V curve in fast evolution}
\end{center}
\end{figure}

Quantum dynamics of MJJs would significantly changes the parity of the Majorana qubits. This gives us an important hint, that is, one could control the voltage behavior of the junction via modulating the Majorana qubit state by a QD. We consider the set-up model shown in Fig. \ref{fig:setup_QD}, where a QD is placed nearby the junction. We choose the simplest model for the QD, which is a single energy level. We place the QD at a very low energy such that there must be one electron occupying on the QD. Thus the system would stay at this state without voltage. If we lift the energy of the QD to a higher energy, the electron on the QD will hop to the MBSs and flip the parity through LZ transition \cite{landau1932b, zener1932}. Then the system would enter the voltage state.

To understand the relation between dynamics of the parity state and the phase difference, we numerically study the system under the linear evolution of the QD energy such as Fig. \ref{fig:I-V curve in fast evolution}(a). The Schr\"{o}dinger equation and the RCSJ equation (\ref{eq:unreduced current}) is self-consistently solved by the fourth-order Runge-Kutta method to represent the features of voltages and hoppings.

\subsection{Fast Evolution}
\begin{figure}[t!]
\begin{center}
	\includegraphics[clip = true, width = 1 \columnwidth]{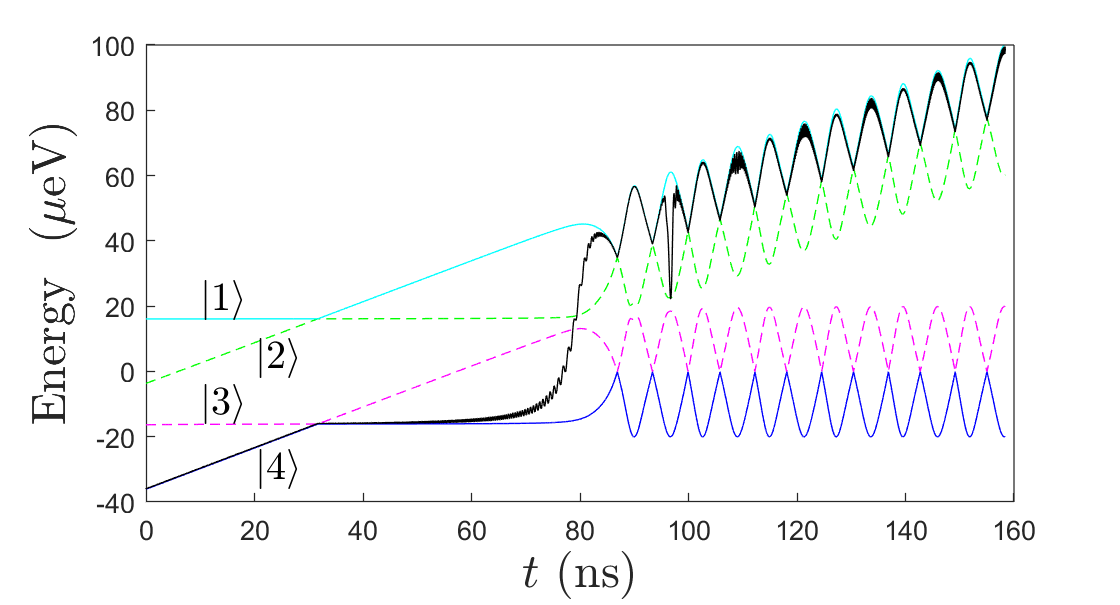}
	\caption{The energy levels (color lines) and weighted occupation curves of levels (black line). We find that the two middle levels have near-crossings at the vicinity of 90 ns. When the QD energy is low, the particle would stay at the lowest level. When the QD energy is high enough, the electron hops from QD to MBSs, however, there is a certain probability to hop back in our model, which is different from the classical tilted-washboard potential picture.}
	\label{fig:energy levels in fast evolution}
\end{center}
\end{figure}

If the system evolves fast, we shall set the total evolution time $t_{\mathrm{tol}}^{\mathrm{fast}}$ equals 158.4 ns approximately. The result of the voltage in Fig. \ref{fig:I-V curve in fast evolution}(b)(g) show a jumping in the voltage curve which is induced by parity flipping and a stable $4\pi$-periodic voltage. Moreover, we observe in Fig. \ref{fig:I-V curve in fast evolution}(c) that there are several peaks decaying with a $4\pi$-period. We want to mention that these peaks should not exist within the classical picture, such as the tilted washboard model mentioned above, since one particle can only occupy one of two potential curves. However, this just can be considered as a possible evidence for the existence of MBSs. We also observe that there is a fast wiggle for the occupation probability curve in Fig. \ref{fig:I-V curve in fast evolution}(c)(e), which is caused by the Larmor precession. Then we show the energy levels (color lines) with a $2\pi$-period and a weighted occupation probability curve (black solid) of energy levels with $4\pi$-periodic negative peaks in Fig. \ref{fig:energy levels in fast evolution}. The Fig. \ref{fig:I-V curve in fast evolution}(f) closely shows us the parity flipping process.

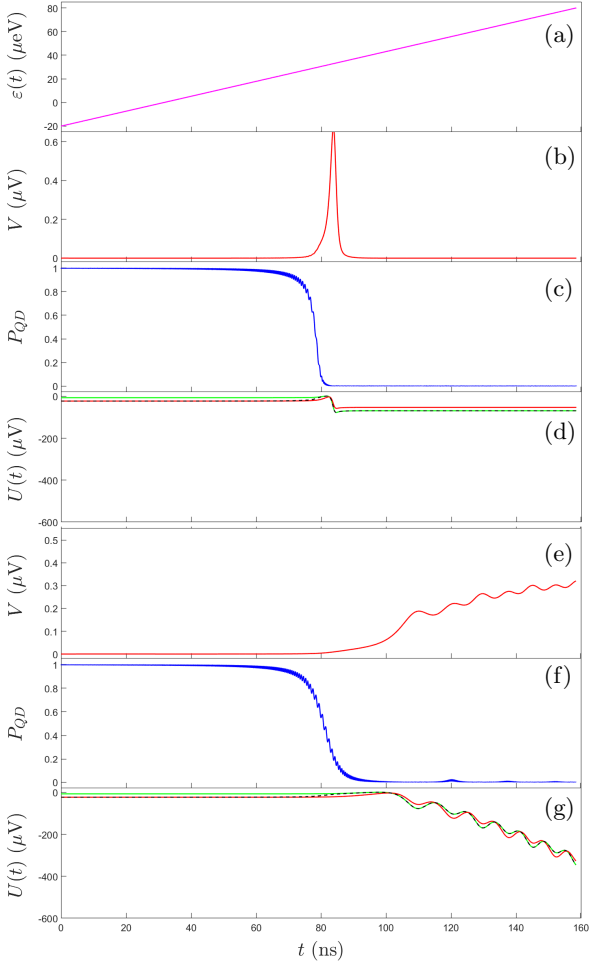
\begin{figure}[t!]
\begin{center}
	\begin{overpic}[clip = true, width = 1 \columnwidth]{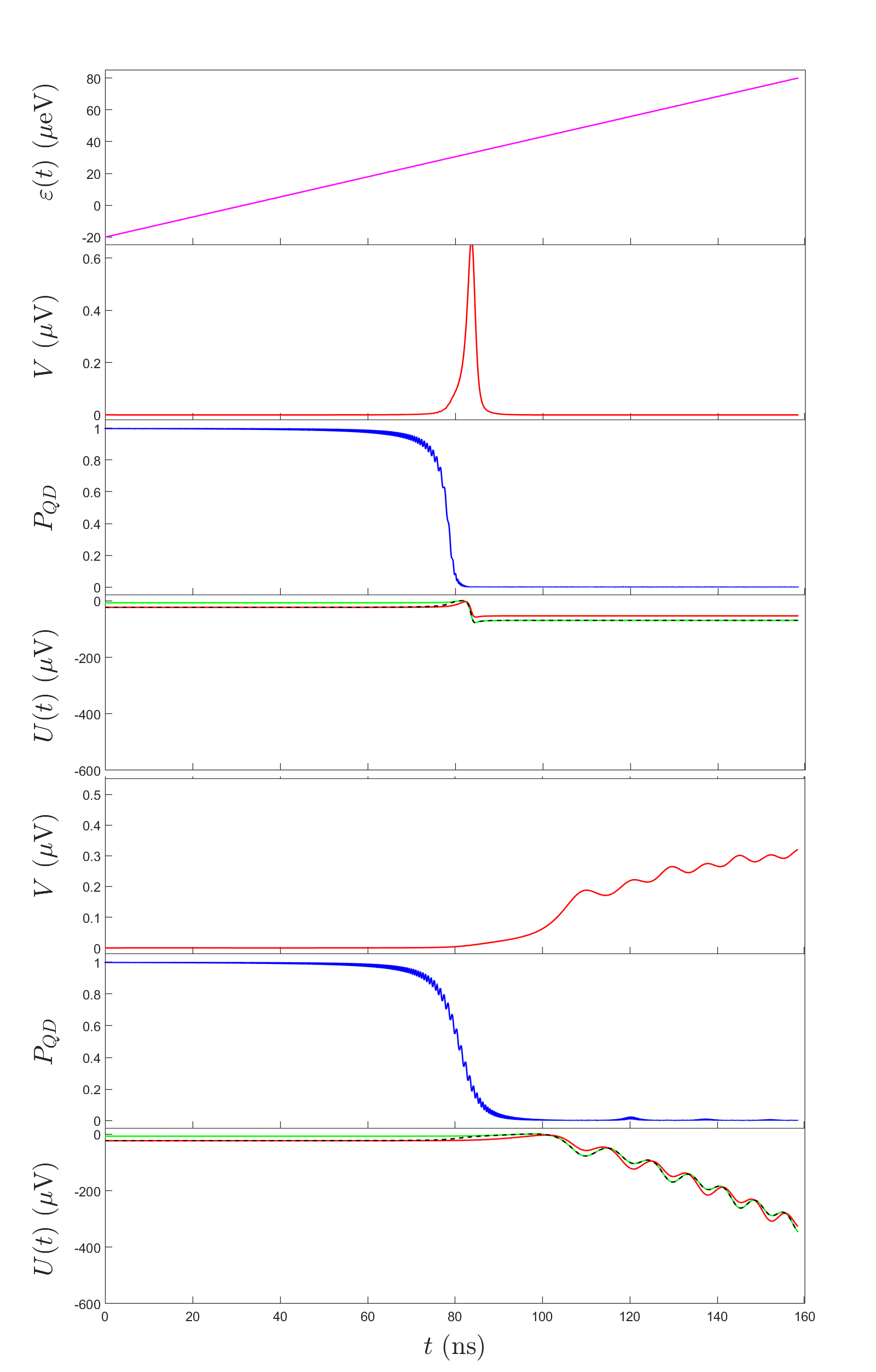}
		\put(55,91){\small (a)}
		\put(55,79){\small (b)}
		\put(55,66){\small (c)}
		\put(55,52){\small (d)}
		\put(55,40){\small (e)}
		\put(55,28){\small (f)}
		\put(55,15){\small (g)}
	\end{overpic}
	\caption{The case of fast evolution where capacitance is changed to (b)$\sim$(d) 0.1, (e)$\sim$(g) 10 multiples of the original value and the others remain unchanged. (a) The energy of the QD. The others are the curves of (b)(e) time-voltage, (c)(f) occupation probability of the QD, and (d)(g) electric potential.}
	\label{fig:I-V curve in fast evolution with other capacitances}
\end{center}
\end{figure}

However, if the capacitance and resistance is adjusted to make the system in overdamped regime with $Q < 1$, there is only a voltage peak with a jumping and falling such as Fig. \ref{fig:I-V curve in fast evolution with other capacitances}(b)$\sim$(d). Else if in lightly damped regime with $Q \gg 1$, the voltage would increase near-linearly and slowly without jumping such as Fig. \ref{fig:I-V curve in fast evolution with other capacitances}(e)$\sim$(g). We neglect these two cases since both of them are short of practical applications.

\subsection{Adiabatic Evolution}
\begin{figure}[t!]
\begin{center}
	\begin{overpic}[clip = true, width = 1 \columnwidth]{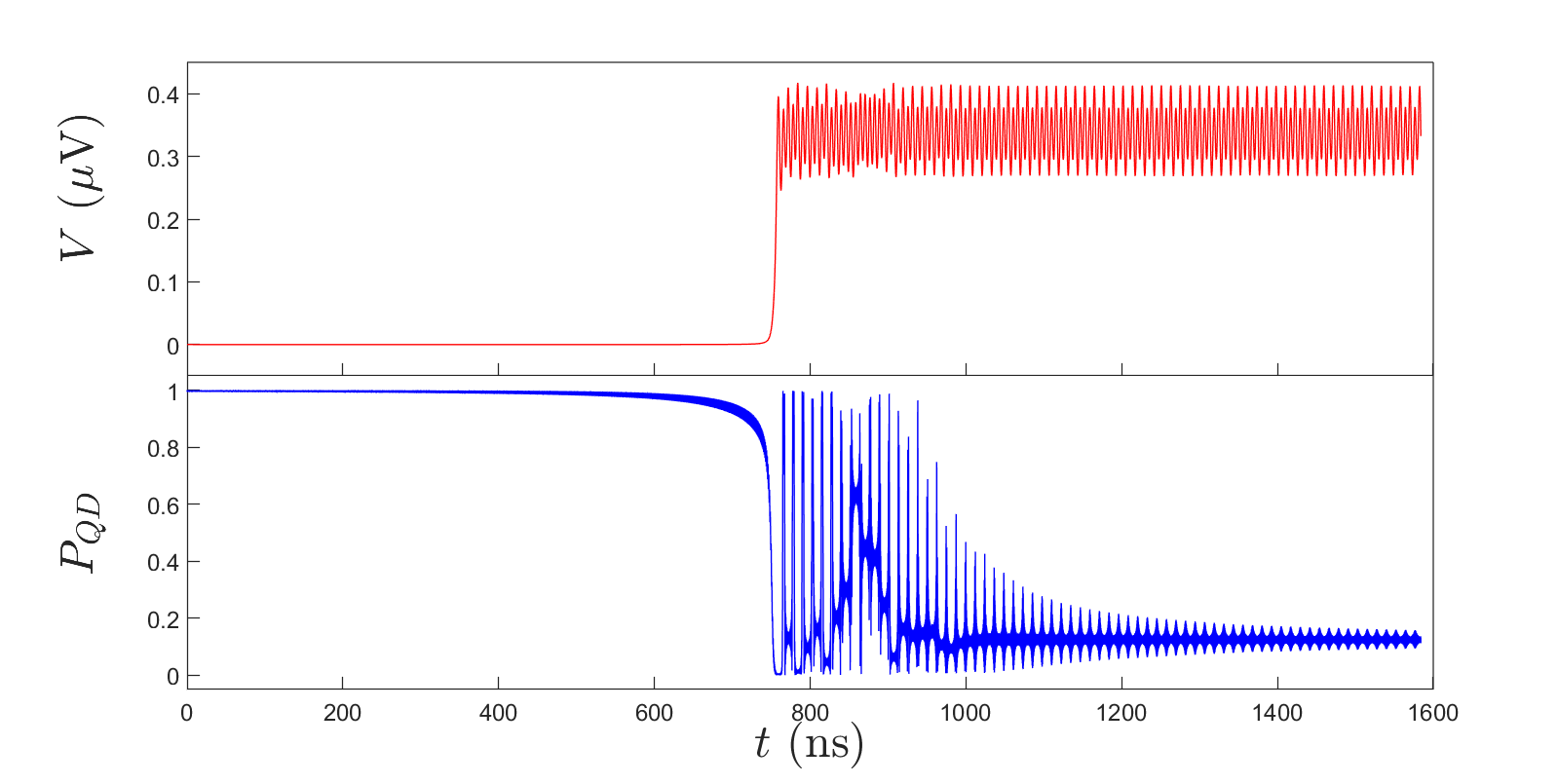}
		\put(15,40){\small (a)}
		\put(15,20){\small (b)}
	\end{overpic}
	\caption{The case of adiabatic evolution. (a) The voltage curve. (b) The occupation probability of the QD with lesser regularity. }
	\label{fig:I-V curve1 in adiabatic evolution}
\end{center}
\end{figure}

Next, if we have the system evolves adiabatically, the hoppings are no longer regular. First, only the evolution time becomes ten times what it was, i.e., the total evolution time $t_{\mathrm{tol}}^{\mathrm{adia}} / t_{\mathrm{tol}}^{\mathrm{fast}} = 10$. Then we observe in Fig. \ref{fig:I-V curve1 in adiabatic evolution}(b) that the occupation probability $P_{\mathrm{QD}}^{}$ still has a $4\pi$-period but more peaks and $P_{\mathrm{QD}}^{}$ does not vanish. By changing the total evolution time, we easily know that the hopping behavior or occupation probability is closely related to the total evolution time.

However, if we change the capacitance $C$ to one hundred times larger and the resistance $R$ to the one-tenth of its original value, the quality factor $Q$ will remain unchanged and the $P_{\mathrm{QD}}^{}$ will behave similarly to Fig. \ref{fig:I-V curve in fast evolution}(c). Not only the hopping peaks but also the voltage is lower than the first two evolution approaches. It is quite easy to verify.

Here, we want point out that several conductance signals of MBSs have been proposed, among which the quantized zero-bias differential conductance peak is most widely adopted \cite{PhysRevLett.115.266804, Kobialka2020681, LIU2020114212, yao_2022}. However, the interpretation of these signals are hindered by the possible existence of low energy Andreev bound states which could induce similar signals \cite{PhysRevB.79.161408}. It has been suggested that MBSs induce a $4\pi$-period supercurrent \cite{PhysRevB.79.161408, PhysRevLett.100.096407}, however, it is susceptible to the quasiparticle poisoning \cite{PhysRevB.94.085409, PhysRevB.85.174533, PhysRevB.86.085414,  PhysRevB.89.140505, PhysRevLett.118.137701}. Hence, our $4\pi$-period result may be considered as a possible evidence for MBSs. So far, we cannot precisely interpret the origin of these peaks.

\subsection{Decoherence by Quasiparticle Poisoning}\label{Section:decoherence}
In fact, the system is coupled to the thermal equilibrium fermionic environment, such that the decoherence from the quasiparticle poisoning should be taken into consideration \cite{RevModPhys.85.623, https://doi.org/10.48550/arxiv.cond-mat/0411174}. From the open quantum system point of view, we can analyse the decoherence problem in this four-level system by introducing the density matrix $\rho(t) = |\Psi(t)\rangle \langle\Psi(t)| = \sum_{i,j=1}^4 \rho_{ij}^{}(t) |i\rangle \langle j|$
and the Lindblad master equation \cite{Breuer2002TheTheoryOfOpenQuantumSystems,nielsen_chuang_2010,PhysRevA.92.012308}
\begin{equation}\label{eq:master equation}
	\dfrac{{\rm d}\rho}{{\rm d}t}
	= -\dfrac{i}{\hbar} [H,\rho]
	+ \sum_{i=1} \dfrac{1}{\tau_i^{}} \left( L_i^{} \rho L_i^\dagger - \dfrac{1}{2} \left\{ L_i^\dagger L_i^{},\rho \right\} \right)
\end{equation}
where $\tau_i^{}$ are the decoherence times and $L_i^{}$ are the all possible standard Lindblad forms:
\begin{align*}
	L_{1,2,3,4}^{} &= |\phi_{1,2,3,4}^{}\rangle \langle\phi_1^{}|, \\
	L_{5,6,7,8}^{} &= |\phi_{1,2,3,4}^{}\rangle \langle\phi_2^{}|, \\
	L_{9,10,11,12}^{} &= |\phi_{1,2,3,4}^{}\rangle \langle\phi_3^{}|, \\
	L_{13,14,15,16}^{} &= |\phi_{1,2,3,4}^{}\rangle \langle\phi_4^{}|,
\end{align*}
with the four instantaneous eigenstates $|\phi_{1,2,3,4}^{}\rangle$ of the Hamiltonian $\mathcal{H}$. Meanwhile, $L_{2,3,4,7,8,12}^{}$ are related to depolarization and then described by depolarization time $\tau_\mathrm{depo}^{}$. And $L_{1,6,11,16}^{}$ are related to dephasing and then described by dephasing time $\tau_\mathrm{deph}^{}$. In general, $\tau_\mathrm{depo}^{} \gg \tau_\mathrm{deph}^{}$ \cite{PhysRevB.85.174533,PhysRevB.86.085414,huang2011observation,hong2013nanoscale,PhysRevB.94.064304}.

\begin{figure}[t!]
\begin{center}
	\begin{overpic}[clip = true, width = 1 \columnwidth]{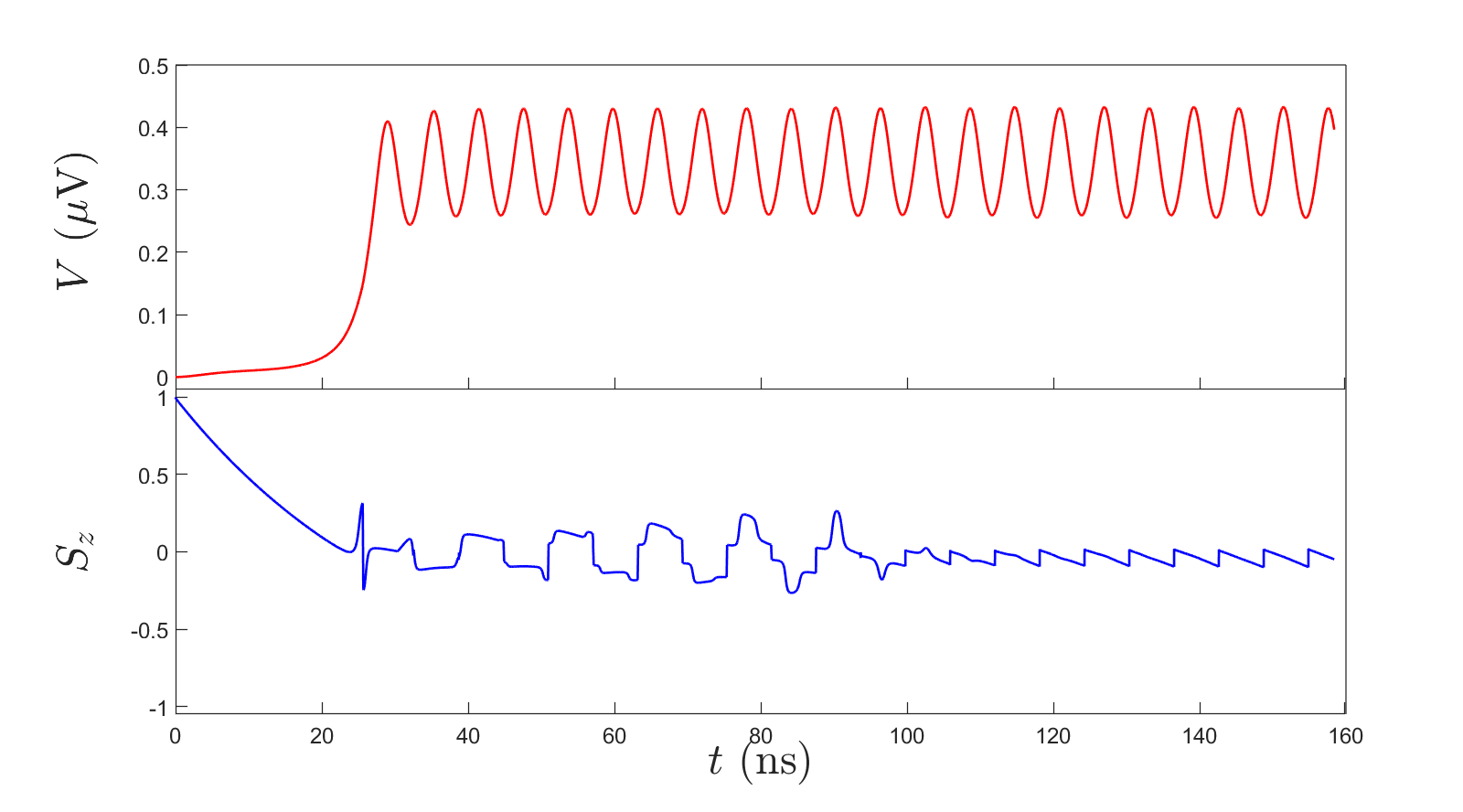}
		\put(85,32){\small (a)}
		\put(85,10){\small (b)}
	\end{overpic}
	\caption{The fast evolution case with decoherence. (a) The induced voltage curve. (b) The average amplitude. }
	\label{fig:decoherence}
\end{center}
\end{figure}

The decoherence time for this process is an exponential function of the superconducting gap \cite{PhysRevB.85.174533,PhysRevB.86.085414},
\begin{equation}
	\dfrac{\hbar}{\tau} = \lambda T e^{-\Delta/T},
\end{equation}
where $\lambda$ is a dimensionless factor estimated around 0.01 and the temperature $T \gg \Delta$ for quasiparticle poisoning processes in nanowire systems \cite{PhysRevB.85.174533}. We present the result of the voltage curve with typical values $\tau_\mathrm{depo}^{} = 1000 \hbar/E_\mathrm{M}^{}$ and $\tau_\mathrm{deph}^{} = 0.1 \hbar/E_\mathrm{M}^{}$ in Fig. \ref{fig:decoherence}. It can be seen that the induced voltage still exists and emerges earlier than the case without the decoherence. By tuning the depolarization and dephasing time, we find that the longer the decoherence times are, the later the induced voltage appears. This agrees with our intuition, because if the decoherence times are long enough, the second term in the right-hand of the master equation (\ref{eq:master equation}) vanishes, and thus the master equation degenerates to Liouville equation. This limiting case exactly corresponds to the previous scenario without the decoherence. Furthermore, the dephasing time has little impact on the induced voltage and is much lesser than it of the depolarization time.

It follows that the induced voltage is robust. Note that the parity does not flip from $1$ to $-1$ any more, but oscillates around zero after the voltage jumping.

\section{Landau-Zener-St\"{u}ckelberg interference}\label{Section:LZS interference}
\begin{figure*}[t!]
\begin{center}
	\begin{overpic}[clip = true, width = 1 \textwidth]{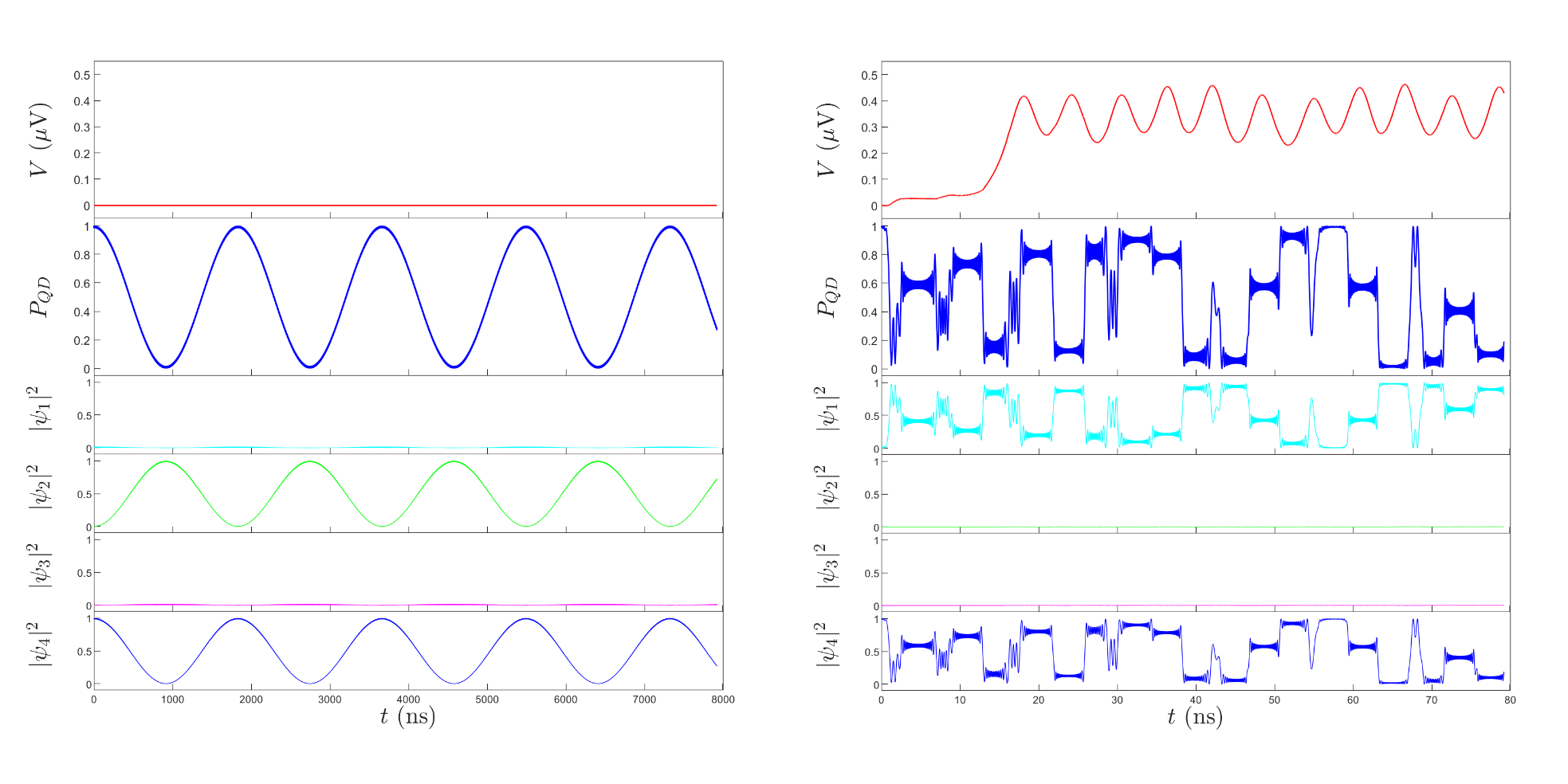}
		\put(43,44){\small (a)}
		\put(43,34){\small (b)}
		\put(43,23){\small (c)}
		\put(43,18){\small (d)}
		\put(43,13){\small (e)}
		\put(43,7){\small (f)}
		\put(93,44){\small (g)}
		\put(93,34){\small (h)}
		\put(93,23){\small (i)}
		\put(93,18){\small (j)}
		\put(93,13){\small (k)}
		\put(93,9){\small (l)}
	\end{overpic}
	\caption{The driving frequency are (a)$\sim$(f) $\omega = 10^{11}$ rad/s and (g)$\sim$(l) $\omega = 10^{8}$ rad/s. (a) No induced voltage. (g) Generating an induced voltage. (b)(h) The pattern of LZS interferences for the QD occupation probability. It can be seen that the pattern in (b) has a fixed period from the LZS interference, but no regularity in (h) since no LZS interference occurs. (c)$\sim$(f), (i)$\sim$(l) are the occupation probabilities of each level.}
	\label{fig:LZS interference}
\end{center}
\end{figure*}

The LZ transition and LZS interference are two kinds of standard quantum behaviors, the latter exhibits a periodic dependence on the phase in a multi-level quantum system in which two of energy levels exhibit avoided level crossings. If a LZ transition is one step in the sequently coherent processes occurring during the dynamics, not only the variation in the occupation probabilities but also the variation in the relative phase between the quantum states is relevant.

Next we will take a two-level system as an example to interpret the LZS interference. Assuming the monochromatic time-dependent bias is
\begin{equation*}
	\varepsilon(t) = \varepsilon_0^{} + A\sin(\omega{t})
\end{equation*}
with the amplitude $A$, driving frequency $\omega$, and offset $\varepsilon_0^{}$. Here we set the degeneracy point as the relative zero. When the ac amplitude $A$ is greater than the dc bias $\varepsilon_0^{}$, the avoided crossing region is passed twice in one ac $2\pi$-cycle, and the LZ transition viewed as a single-passage process may occur every time.
We can experimentally control the driving frequency $\omega$ so that the time interval between two adjacent avoided crossings, i.e., half driving period $T/2 = \pi/\omega$, is much less than the dephasing time of the system $\tau_{\mathrm{deph}}^{}$ and larger than the characteristic time for a LZ transition $t_{\mathrm{LZ}}^{}$. Thus, if the system starts at one of levels, it will have a certain population at both levels after the first LZ transition. Then the wave functions of the two levels evolve separately with different energy eigenvalues.
In the case of double-passage, the phase accumulated during the evolution process between two LZ transitions, of the upper level is unequal to it of the lower level. The sum of phase change of the upper and lower wave functions during this process is called ``St\"{u}ckelberg phase", $\Phi^{\mathrm{St}}$, which includes two parts: the first $\varphi_{1,2}^{\mathrm{adia}}$ is the adiabatic phase from the adiabatic evolution process away from the degeneracy point, where the subscripts $1,2$ respectively represent the first and second half-cycles correspondingly; the second $\widetilde{\varphi}^{\mathrm{S}}$ is the diabatic phase from the non-adiabatic LZ transitions in the vicinity of the degeneracy point, which equals to Stokes phase $\varphi^{\mathrm{S}}$ minus $\pi/2$ \cite{SHEVCHENKO20101, IVAKHNENKO20231, 10.1360/132012-930}.
When the system passes through the degeneracy point second time in a cycle, the wave functions of different levels are still coherent because the time difference is within the dephasing time, and the interference occurs at this point. This interference is called ``Landau-Zener-St\"{u}ckelberg interference" \cite{landau1932b, zener1932, stuckelberg1932theory}.

When the phase difference between the two wave functions equals $2n\pi$, or the adiabatic phase $\varphi_{1,2}^{\mathrm{adia}}$ equals $\pi/2 + n\pi$, the constructive interference occurs and the corresponding LZ transition probability is at the maximum value. Similarly, when the phase difference between the two wave functions equals $(2n+1)\pi$, or the adiabatic phase $\varphi_{1,2}^{\mathrm{adia}}$ equals $n\pi$, the destructive interference occurs and the corresponding LZ transition probability is at the minimum value.

Now we study the LZS interferences occurring in our four-level MJJ. Compared to the order of the other times, the dephasing time $\tau_{\mathrm{deph}}^{}$ here can be viewed as a large number. For simplicity, we set the offset $\varepsilon_0^{} = 0$, that is, the time-dependent QD energy level equals
\begin{equation}
	\varepsilon(t) = A\sin(\omega{t})
\end{equation}
where the amplitude $A / E_\mathrm{M}^{} = 2$. Thus, the Hamiltonian becomes
\begin{widetext}
\begin{equation}\label{eq:LZS interference Hamiltonian}
	\mathcal{H}_{\mathrm{LZS}}^{} = 
	\begin{pmatrix}
		E_\mathrm{M}^{}\cos{\frac{\theta}{2}} & \delta_1^{} & 0 & T \\
		\delta_1^{} & -E_\mathrm{M}^{}\cos{\frac{\theta}{2}} & -T & 0 \\
		0 & -T & E_\mathrm{M}^{}\cos{\frac{\theta}{2}} + A\sin(\omega{t}) & \delta_2^{} \\
		T & 0 & \delta_2^{} & -E_\mathrm{M}^{}\cos{\frac{\theta}{2} + A\sin(\omega{t})}
	\end{pmatrix}.
\end{equation}
\end{widetext}
Fig. \ref{fig:LZS interference} shows two useful results under two special values. When $\omega$ is small enough ($\sim 10$ rad/s), the voltage will not be induced. In this process, the electron only occupies the levels $|2\rangle$ and $|4\rangle$, so there is no parity flipping occurring. On the other hand, while $\omega$ is large enough ($\sim 10^{29}$ rad/s), the electron only occupies the levels $|1\rangle$ and $|4\rangle$ (unless the evolution time is very long), so that the voltage will be induced with a sequence of parity flipping. The other cases are hard to uniformly describe because of their complication and irregulation. For simplicity, we call these irregular cases as intermediate region. When driving frequency runs from $\sim 10$ rad/s to $\sim 10^{29}$ rad/s, the voltage-free, intermediate and voltage regions would appear in turn. Thus, we can tune the driving frequency $\omega$ to realize the switch between the voltage and voltage-free states.


\section{Damped Rabi oscillation}
\label{Section:Rabi oscillation}
\begin{figure}[t!]
\begin{center}
	\begin{overpic}[clip = true, width = 1 \columnwidth]{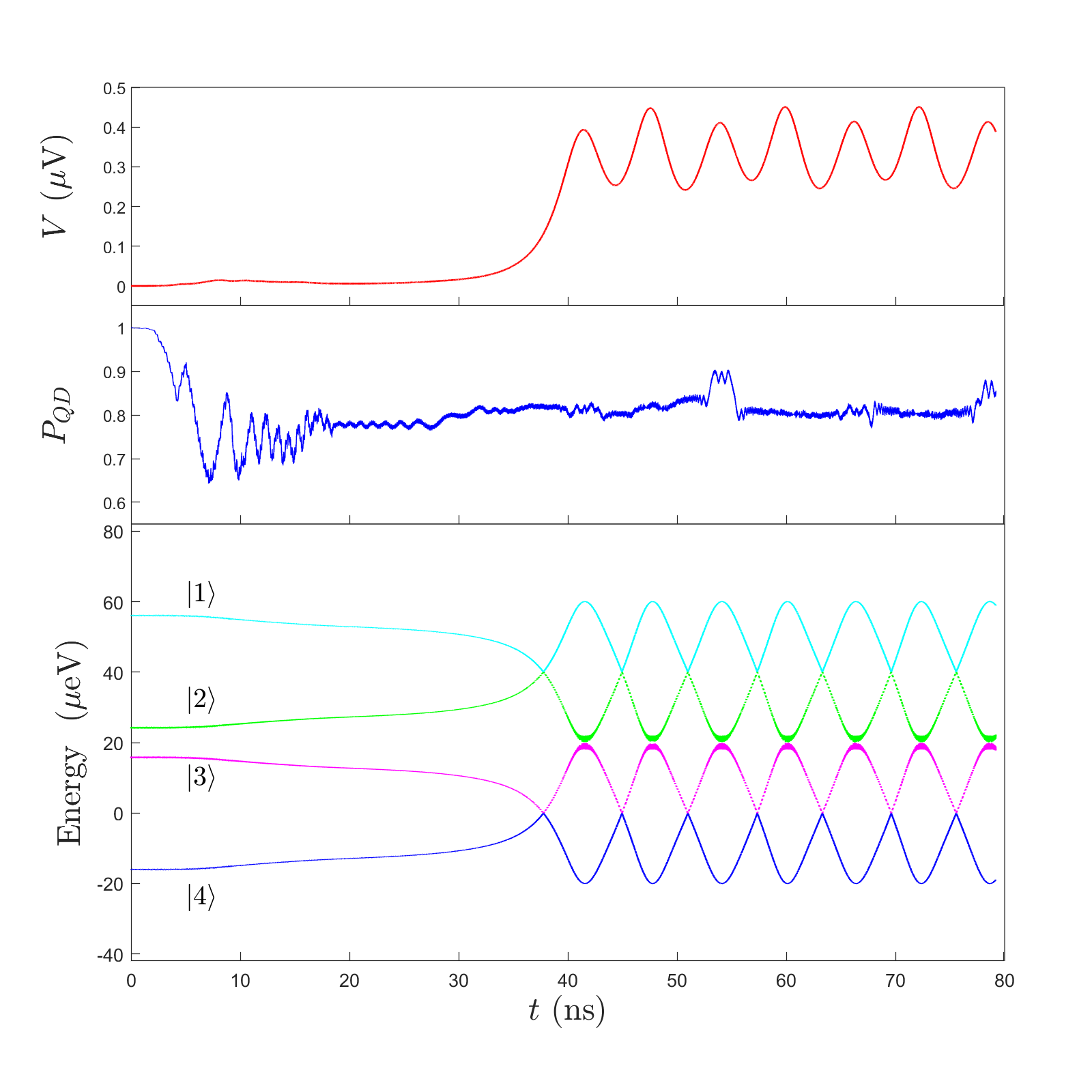}
		\put(85,88){\small (a)}
		\put(85,68){\small (b)}
		\put(85,48){\small (c)}
	\end{overpic}
	\caption{(a) The induced voltage. (b) The damped Rabi oscillation in the QD occupation probability. (c) The energy levels with crossings.}
	\label{fig:Rabi oscillations}
\end{center}
\end{figure}

We all know that the Rabi oscillation, arised from the perturbation caused by the anti-diagonal matrix elements, shall occur in a multi-level system driven by an alternating force \cite{PhysRev.51.652,griffiths2018introduction}. Assuming the time-dependent tunneling coefficient $T(t)$ equals
\begin{equation}
	T(t) = T_0^{} \cos(\omega{t})
\end{equation}
where $T_0^{} / E_\mathrm{M}^{} = 0.01$, $\omega = 2 \varepsilon_0^{}/\hbar$. Note that the energy of the QD becomes a fixed value $\varepsilon_0^{} = 2 E_\mathrm{M}^{}$. The Hamiltonian changes to
\begin{align}\label{eq:Rabi oscillation Hamiltonian}
	& \mathcal{H}_{\mathrm{Rabi}}^{} = \notag\\
	& \begin{pmatrix}
		E_\mathrm{M}^{}\cos{\frac{\theta}{2}} & \delta_1^{} & 0 & T_0^{} \cos(\omega{t}) \\
		\delta_1^{} & -E_\mathrm{M}^{}\cos{\frac{\theta}{2}} & -T_0^{} \cos(\omega{t}) & 0 \\
		0 & -T_0^{} \cos(\omega{t}) & E_\mathrm{M}^{}\cos{\frac{\theta}{2}}+\varepsilon_0^{} & \delta_2^{} \\
		T_0^{} \cos(\omega{t}) & 0 & \delta_2^{} & -E_\mathrm{M}^{}\cos{\frac{\theta}{2}+\varepsilon_0^{}}
	\end{pmatrix},
\end{align}
and the result is shown in Fig. \ref{fig:Rabi oscillations}.

We pay attention to the first 30 ns since the Rabi oscillation does not dominate the subsequent dynamics during the period. The rest of the process is driven by a combination of several mechanisms, so there is no regularity after the voltage jumping and level-crossings appearing simultaneously. We can see that the Rabi oscillation is suppressed by a damping which is caused by the coupling of nonlinearity and qubits, whose behavior is similar to the related theories and experiments \cite{PhysRevB.72.172509, doi:10.1063/1.5078628, Schon2020, Shevchenko_2008, PhysRevLett.95.067001, PhysRevB.78.104510}. Our results have important implication for quantum state engineering of Josephson devices and quantum computation based on Majorana qubits \cite{doi:10.1126/science.1069452}.

\section{Conclusions and Discussions}\label{Section:Conclusions}
In this work, we utilize a formalism to study our MJJ set-up with a reliable RCSJ model. Then an internal $4\pi$-periodic hopping behavior, which comes from quantum dynamics of the Majorana parity state, is found in the QD occupation probability curve. As shown numerically, these hopping peaks could have similar curves for the same quality factor and different total evolution times, and is remarkably different from the classical mechanical analog picture. Thus this hopping signal can be used to confirm the existence of MBSs. From the perspective of energy level, the LZ transition in this four-level system is a direct transition from lowest level to highest level, which is different from LZ transition caused by level avoided crossings in two-level system. Besides, the easily measurable $4\pi$-periodic voltage signal induced by parity flipping can also be treated as a possible evidence for the existence of MBSs. We consider the effect of decoherence and successfully check the robustness of this $4\pi$-periodic voltage.

Then, we investigate the Landau-Zener-St\"{u}ckelberg interference by changing the QD energy evolution approach. We also realize the switch between voltage and voltage-free states by tuning the driving frequency.

Next, when tunneling quantity evolves cosinoidally, we observe that the set-up generates damped Rabi oscillation in the QD occupation probability curve and its amplitude decays with time before the voltage jumping. Rabi oscillation is an important phenomenon in quantum circuits. We find that the form and order of the damped Rabi oscillation are similar to the existing experimental results. It gives us confidence to manipulate qubits by MBSs and greatly enhances the prospects of quantum computation with Majorana qubits.

Last, based on the fact that the voltage-free and voltage states consist a two-state classical system which corresponds to the low (binary $0$) and high (binary $1$) level output voltage in classical circuits, and has a one-to-one correspondence between it and original two-state quantum system, (specifically, two quantum states, in the case of LZ transition, are the occupied ground state $|4\rangle$ and the occupied third excited state $|1\rangle$; in the case of LZS interference, are the combination of $|1\rangle$ and $|4\rangle$ and the combination of $|2\rangle$ and $|4\rangle$,) this set-up might be used to convert Majorana qubits to classical bits.

In conclusion, we have studied a MJJ set-up with a QD and found rich dynamical response. The behaviors and signals mentioned above are easy to be measured experimentally so that can be utilized to determine whether or not MBSs exist possibly. On the other hand, our set-up can be used to transfer information between Majorana qubits and classical bits, which might help us design circuit elements for hybrid quantum-classical computers.

\section*{Acknowledgements}
This project is supported by NKRDPC-2018YFA0305603, NKRDPC-2022YFA1402802, NSFC-12174453, NSFC-92165204, NSFC-11974432, GBABRF-2019A1515011620, Shenzhen Institute for Quantum Science and Engineering (Grant No. SIQSE202102), and Leading Talent Program of Guangdong Special Projects (201626003).

We would like to thank Jun Li for examining Fig. \ref{fig:I-V curve in fast evolution}(c), and thank Jia-Jin Feng, Jia-Zheng Ma, Yun-Feng Chen, Biao Lyu and Zenan Liu for valuable discussions. 

\bibliographystyle{apsrev4-2}
\bibliography{ReferencesOfParityFlipping}

\begin{thebibliography}{67}%
\makeatletter
\providecommand \@ifxundefined [1]{%
 \@ifx{#1\undefined}
}%
\providecommand \@ifnum [1]{%
 \ifnum #1\expandafter \@firstoftwo
 \else \expandafter \@secondoftwo
 \fi
}%
\providecommand \@ifx [1]{%
 \ifx #1\expandafter \@firstoftwo
 \else \expandafter \@secondoftwo
 \fi
}%
\providecommand \natexlab [1]{#1}%
\providecommand \enquote  [1]{``#1''}%
\providecommand \bibnamefont  [1]{#1}%
\providecommand \bibfnamefont [1]{#1}%
\providecommand \citenamefont [1]{#1}%
\providecommand \href@noop [0]{\@secondoftwo}%
\providecommand \href [0]{\begingroup \@sanitize@url \@href}%
\providecommand \@href[1]{\@@startlink{#1}\@@href}%
\providecommand \@@href[1]{\endgroup#1\@@endlink}%
\providecommand \@sanitize@url [0]{\catcode `\\12\catcode `\$12\catcode
  `\&12\catcode `\#12\catcode `\^12\catcode `\_12\catcode `\%12\relax}%
\providecommand \@@startlink[1]{}%
\providecommand \@@endlink[0]{}%
\providecommand \url  [0]{\begingroup\@sanitize@url \@url }%
\providecommand \@url [1]{\endgroup\@href {#1}{\urlprefix }}%
\providecommand \urlprefix  [0]{URL }%
\providecommand \Eprint [0]{\href }%
\providecommand \doibase [0]{https://doi.org/}%
\providecommand \selectlanguage [0]{\@gobble}%
\providecommand \bibinfo  [0]{\@secondoftwo}%
\providecommand \bibfield  [0]{\@secondoftwo}%
\providecommand \translation [1]{[#1]}%
\providecommand \BibitemOpen [0]{}%
\providecommand \bibitemStop [0]{}%
\providecommand \bibitemNoStop [0]{.\EOS\space}%
\providecommand \EOS [0]{\spacefactor3000\relax}%
\providecommand \BibitemShut  [1]{\csname bibitem#1\endcsname}%
\let\auto@bib@innerbib\@empty
\bibitem [{\citenamefont {Albrecht}\ \emph {et~al.}(2016)\citenamefont
  {Albrecht}, \citenamefont {Higginbotham}, \citenamefont {Madsen},
  \citenamefont {Kuemmeth}, \citenamefont {Jespersen}, \citenamefont
  {Nyg{\aa}rd}, \citenamefont {Krogstrup},\ and\ \citenamefont
  {Marcus}}]{albrecht2016exponential}%
  \BibitemOpen
  \bibfield  {author} {\bibinfo {author} {\bibfnamefont {S.~M.}\ \bibnamefont
  {Albrecht}}, \bibinfo {author} {\bibfnamefont {A.~P.}\ \bibnamefont
  {Higginbotham}}, \bibinfo {author} {\bibfnamefont {M.}~\bibnamefont
  {Madsen}}, \bibinfo {author} {\bibfnamefont {F.}~\bibnamefont {Kuemmeth}},
  \bibinfo {author} {\bibfnamefont {T.~S.}\ \bibnamefont {Jespersen}}, \bibinfo
  {author} {\bibfnamefont {J.}~\bibnamefont {Nyg{\aa}rd}}, \bibinfo {author}
  {\bibfnamefont {P.}~\bibnamefont {Krogstrup}},\ and\ \bibinfo {author}
  {\bibfnamefont {C.}~\bibnamefont {Marcus}},\ }\href
  {https://doi.org/https://doi.org/10.1038/531177a} {\bibfield  {journal}
  {\bibinfo  {journal} {Nature}\ }\textbf {\bibinfo {volume} {531}},\ \bibinfo
  {pages} {206} (\bibinfo {year} {2016})}\BibitemShut {NoStop}%
\bibitem [{\citenamefont {Sato}\ and\ \citenamefont
  {Fujimoto}(2016)}]{doi:10.7566/JPSJ.85.072001}%
  \BibitemOpen
  \bibfield  {author} {\bibinfo {author} {\bibfnamefont {M.}~\bibnamefont
  {Sato}}\ and\ \bibinfo {author} {\bibfnamefont {S.}~\bibnamefont
  {Fujimoto}},\ }\href {https://doi.org/10.7566/JPSJ.85.072001} {\bibfield
  {journal} {\bibinfo  {journal} {Journal of the Physical Society of Japan}\
  }\textbf {\bibinfo {volume} {85}},\ \bibinfo {pages} {072001} (\bibinfo
  {year} {2016})}\BibitemShut {NoStop}%
\bibitem [{\citenamefont {Zhu}\ \emph {et~al.}(2017)\citenamefont {Zhu},
  \citenamefont {Wang},\ and\ \citenamefont {Zhang}}]{Zhu_2016}%
  \BibitemOpen
  \bibfield  {author} {\bibinfo {author} {\bibfnamefont {G.-Y.}\ \bibnamefont
  {Zhu}}, \bibinfo {author} {\bibfnamefont {R.-R.}\ \bibnamefont {Wang}},\ and\
  \bibinfo {author} {\bibfnamefont {G.-M.}\ \bibnamefont {Zhang}},\ }\href
  {https://doi.org/10.7693/wl20170303} {\bibfield  {journal} {\bibinfo
  {journal} {WuLi}\ }\textbf {\bibinfo {volume} {46}},\ \bibinfo {pages} {154}
  (\bibinfo {year} {2017})}\BibitemShut {NoStop}%
\bibitem [{\citenamefont {Liang}\ \emph {et~al.}(2020)\citenamefont {Liang},
  \citenamefont {Wang}, \citenamefont {Kawakami},\ and\ \citenamefont
  {Hu}}]{Wang_2020}%
  \BibitemOpen
  \bibfield  {author} {\bibinfo {author} {\bibfnamefont {Q.-F.}\ \bibnamefont
  {Liang}}, \bibinfo {author} {\bibfnamefont {Z.}~\bibnamefont {Wang}},
  \bibinfo {author} {\bibfnamefont {T.}~\bibnamefont {Kawakami}},\ and\
  \bibinfo {author} {\bibfnamefont {X.}~\bibnamefont {Hu}},\ }\href
  {https://doi.org/10.7498/aps.69.20190959} {\bibfield  {journal} {\bibinfo
  {journal} {Acta Phys. Sin.}\ }\textbf {\bibinfo {volume} {69}},\ \bibinfo
  {pages} {117102} (\bibinfo {year} {2020})}\BibitemShut {NoStop}%
\bibitem [{\citenamefont {Kitaev}(2003)}]{KITAEV20032}%
  \BibitemOpen
  \bibfield  {author} {\bibinfo {author} {\bibfnamefont {A.}~\bibnamefont
  {Kitaev}},\ }\href
  {https://doi.org/https://doi.org/10.1016/S0003-4916(02)00018-0} {\bibfield
  {journal} {\bibinfo  {journal} {Annals of Physics}\ }\textbf {\bibinfo
  {volume} {303}},\ \bibinfo {pages} {2} (\bibinfo {year} {2003})}\BibitemShut
  {NoStop}%
\bibitem [{\citenamefont {Nayak}\ \emph {et~al.}(2008)\citenamefont {Nayak},
  \citenamefont {Simon}, \citenamefont {Stern}, \citenamefont {Freedman},\ and\
  \citenamefont {Das~Sarma}}]{RevModPhys.80.1083}%
  \BibitemOpen
  \bibfield  {author} {\bibinfo {author} {\bibfnamefont {C.}~\bibnamefont
  {Nayak}}, \bibinfo {author} {\bibfnamefont {S.~H.}\ \bibnamefont {Simon}},
  \bibinfo {author} {\bibfnamefont {A.}~\bibnamefont {Stern}}, \bibinfo
  {author} {\bibfnamefont {M.}~\bibnamefont {Freedman}},\ and\ \bibinfo
  {author} {\bibfnamefont {S.}~\bibnamefont {Das~Sarma}},\ }\href
  {https://doi.org/10.1103/RevModPhys.80.1083} {\bibfield  {journal} {\bibinfo
  {journal} {Rev. Mod. Phys.}\ }\textbf {\bibinfo {volume} {80}},\ \bibinfo
  {pages} {1083} (\bibinfo {year} {2008})}\BibitemShut {NoStop}%
\bibitem [{\citenamefont {Field}\ and\ \citenamefont
  {Simula}(2018)}]{Field_2018}%
  \BibitemOpen
  \bibfield  {author} {\bibinfo {author} {\bibfnamefont {B.}~\bibnamefont
  {Field}}\ and\ \bibinfo {author} {\bibfnamefont {T.}~\bibnamefont {Simula}},\
  }\href {https://doi.org/10.1088/2058-9565/aacad2} {\bibfield  {journal}
  {\bibinfo  {journal} {Quantum Science and Technology}\ }\textbf {\bibinfo
  {volume} {3}},\ \bibinfo {pages} {045004} (\bibinfo {year}
  {2018})}\BibitemShut {NoStop}%
\bibitem [{\citenamefont {Makhlin}\ \emph {et~al.}(2001)\citenamefont
  {Makhlin}, \citenamefont {Sch\"on},\ and\ \citenamefont
  {Shnirman}}]{RevModPhys.73.357}%
  \BibitemOpen
  \bibfield  {author} {\bibinfo {author} {\bibfnamefont {Y.}~\bibnamefont
  {Makhlin}}, \bibinfo {author} {\bibfnamefont {G.}~\bibnamefont {Sch\"on}},\
  and\ \bibinfo {author} {\bibfnamefont {A.}~\bibnamefont {Shnirman}},\ }\href
  {https://doi.org/10.1103/RevModPhys.73.357} {\bibfield  {journal} {\bibinfo
  {journal} {Rev. Mod. Phys.}\ }\textbf {\bibinfo {volume} {73}},\ \bibinfo
  {pages} {357} (\bibinfo {year} {2001})}\BibitemShut {NoStop}%
\bibitem [{\citenamefont {Sato}\ \emph {et~al.}(2009)\citenamefont {Sato},
  \citenamefont {Takahashi},\ and\ \citenamefont
  {Fujimoto}}]{PhysRevLett.103.020401}%
  \BibitemOpen
  \bibfield  {author} {\bibinfo {author} {\bibfnamefont {M.}~\bibnamefont
  {Sato}}, \bibinfo {author} {\bibfnamefont {Y.}~\bibnamefont {Takahashi}},\
  and\ \bibinfo {author} {\bibfnamefont {S.}~\bibnamefont {Fujimoto}},\ }\href
  {https://doi.org/10.1103/PhysRevLett.103.020401} {\bibfield  {journal}
  {\bibinfo  {journal} {Phys. Rev. Lett.}\ }\textbf {\bibinfo {volume} {103}},\
  \bibinfo {pages} {020401} (\bibinfo {year} {2009})}\BibitemShut {NoStop}%
\bibitem [{\citenamefont {Song}\ \emph {et~al.}(2022)\citenamefont {Song},
  \citenamefont {Zhang},\ and\ \citenamefont {Hao}}]{PhysRevLett.128.016402}%
  \BibitemOpen
  \bibfield  {author} {\bibinfo {author} {\bibfnamefont {R.}~\bibnamefont
  {Song}}, \bibinfo {author} {\bibfnamefont {P.}~\bibnamefont {Zhang}},\ and\
  \bibinfo {author} {\bibfnamefont {N.}~\bibnamefont {Hao}},\ }\href
  {https://doi.org/10.1103/PhysRevLett.128.016402} {\bibfield  {journal}
  {\bibinfo  {journal} {Phys. Rev. Lett.}\ }\textbf {\bibinfo {volume} {128}},\
  \bibinfo {pages} {016402} (\bibinfo {year} {2022})}\BibitemShut {NoStop}%
\bibitem [{\citenamefont {Fu}\ and\ \citenamefont
  {Kane}(2008)}]{PhysRevLett.100.096407}%
  \BibitemOpen
  \bibfield  {author} {\bibinfo {author} {\bibfnamefont {L.}~\bibnamefont
  {Fu}}\ and\ \bibinfo {author} {\bibfnamefont {C.~L.}\ \bibnamefont {Kane}},\
  }\href {https://doi.org/10.1103/PhysRevLett.100.096407} {\bibfield  {journal}
  {\bibinfo  {journal} {Phys. Rev. Lett.}\ }\textbf {\bibinfo {volume} {100}},\
  \bibinfo {pages} {096407} (\bibinfo {year} {2008})}\BibitemShut {NoStop}%
\bibitem [{\citenamefont {Qi}\ and\ \citenamefont
  {Zhang}(2011)}]{RevModPhys.83.1057}%
  \BibitemOpen
  \bibfield  {author} {\bibinfo {author} {\bibfnamefont {X.-L.}\ \bibnamefont
  {Qi}}\ and\ \bibinfo {author} {\bibfnamefont {S.-C.}\ \bibnamefont {Zhang}},\
  }\href {https://doi.org/10.1103/RevModPhys.83.1057} {\bibfield  {journal}
  {\bibinfo  {journal} {Rev. Mod. Phys.}\ }\textbf {\bibinfo {volume} {83}},\
  \bibinfo {pages} {1057} (\bibinfo {year} {2011})}\BibitemShut {NoStop}%
\bibitem [{\citenamefont {Read}\ and\ \citenamefont
  {Green}(2000)}]{PhysRevB.61.10267}%
  \BibitemOpen
  \bibfield  {author} {\bibinfo {author} {\bibfnamefont {N.}~\bibnamefont
  {Read}}\ and\ \bibinfo {author} {\bibfnamefont {D.}~\bibnamefont {Green}},\
  }\href {https://doi.org/10.1103/PhysRevB.61.10267} {\bibfield  {journal}
  {\bibinfo  {journal} {Phys. Rev. B}\ }\textbf {\bibinfo {volume} {61}},\
  \bibinfo {pages} {10267} (\bibinfo {year} {2000})}\BibitemShut {NoStop}%
\bibitem [{\citenamefont {Lutchyn}\ \emph {et~al.}(2010)\citenamefont
  {Lutchyn}, \citenamefont {Sau},\ and\ \citenamefont
  {Das~Sarma}}]{PhysRevLett.105.077001}%
  \BibitemOpen
  \bibfield  {author} {\bibinfo {author} {\bibfnamefont {R.~M.}\ \bibnamefont
  {Lutchyn}}, \bibinfo {author} {\bibfnamefont {J.~D.}\ \bibnamefont {Sau}},\
  and\ \bibinfo {author} {\bibfnamefont {S.}~\bibnamefont {Das~Sarma}},\ }\href
  {https://doi.org/10.1103/PhysRevLett.105.077001} {\bibfield  {journal}
  {\bibinfo  {journal} {Phys. Rev. Lett.}\ }\textbf {\bibinfo {volume} {105}},\
  \bibinfo {pages} {077001} (\bibinfo {year} {2010})}\BibitemShut {NoStop}%
\bibitem [{\citenamefont {Mourik}\ \emph {et~al.}(2012)\citenamefont {Mourik},
  \citenamefont {Zuo}, \citenamefont {Frolov}, \citenamefont {Plissard},
  \citenamefont {Bakkers},\ and\ \citenamefont
  {Kouwenhoven}}]{doi:10.1126/science.1222360}%
  \BibitemOpen
  \bibfield  {author} {\bibinfo {author} {\bibfnamefont {V.}~\bibnamefont
  {Mourik}}, \bibinfo {author} {\bibfnamefont {K.}~\bibnamefont {Zuo}},
  \bibinfo {author} {\bibfnamefont {S.~M.}\ \bibnamefont {Frolov}}, \bibinfo
  {author} {\bibfnamefont {S.~R.}\ \bibnamefont {Plissard}}, \bibinfo {author}
  {\bibfnamefont {E.~P. A.~M.}\ \bibnamefont {Bakkers}},\ and\ \bibinfo
  {author} {\bibfnamefont {L.~P.}\ \bibnamefont {Kouwenhoven}},\ }\href
  {https://doi.org/10.1126/science.1222360} {\bibfield  {journal} {\bibinfo
  {journal} {Science}\ }\textbf {\bibinfo {volume} {336}},\ \bibinfo {pages}
  {1003} (\bibinfo {year} {2012})}\BibitemShut {NoStop}%
\bibitem [{\citenamefont {Xu}\ \emph {et~al.}(2015)\citenamefont {Xu},
  \citenamefont {Wang}, \citenamefont {Liu}, \citenamefont {Ge}, \citenamefont
  {Yang}, \citenamefont {Liu}, \citenamefont {Xu}, \citenamefont {Guan},
  \citenamefont {Gao}, \citenamefont {Qian}, \citenamefont {Liu}, \citenamefont
  {Wang}, \citenamefont {Zhang}, \citenamefont {Xue},\ and\ \citenamefont
  {Jia}}]{PhysRevLett.114.017001}%
  \BibitemOpen
  \bibfield  {author} {\bibinfo {author} {\bibfnamefont {J.-P.}\ \bibnamefont
  {Xu}}, \bibinfo {author} {\bibfnamefont {M.-X.}\ \bibnamefont {Wang}},
  \bibinfo {author} {\bibfnamefont {Z.~L.}\ \bibnamefont {Liu}}, \bibinfo
  {author} {\bibfnamefont {J.-F.}\ \bibnamefont {Ge}}, \bibinfo {author}
  {\bibfnamefont {X.}~\bibnamefont {Yang}}, \bibinfo {author} {\bibfnamefont
  {C.}~\bibnamefont {Liu}}, \bibinfo {author} {\bibfnamefont {Z.~A.}\
  \bibnamefont {Xu}}, \bibinfo {author} {\bibfnamefont {D.}~\bibnamefont
  {Guan}}, \bibinfo {author} {\bibfnamefont {C.~L.}\ \bibnamefont {Gao}},
  \bibinfo {author} {\bibfnamefont {D.}~\bibnamefont {Qian}}, \bibinfo {author}
  {\bibfnamefont {Y.}~\bibnamefont {Liu}}, \bibinfo {author} {\bibfnamefont
  {Q.-H.}\ \bibnamefont {Wang}}, \bibinfo {author} {\bibfnamefont {F.-C.}\
  \bibnamefont {Zhang}}, \bibinfo {author} {\bibfnamefont {Q.-K.}\ \bibnamefont
  {Xue}},\ and\ \bibinfo {author} {\bibfnamefont {J.-F.}\ \bibnamefont {Jia}},\
  }\href {https://doi.org/10.1103/PhysRevLett.114.017001} {\bibfield  {journal}
  {\bibinfo  {journal} {Phys. Rev. Lett.}\ }\textbf {\bibinfo {volume} {114}},\
  \bibinfo {pages} {017001} (\bibinfo {year} {2015})}\BibitemShut {NoStop}%
\bibitem [{\citenamefont {Alicea}(2012)}]{Alicea_2012}%
  \BibitemOpen
  \bibfield  {author} {\bibinfo {author} {\bibfnamefont {J.}~\bibnamefont
  {Alicea}},\ }\href {https://doi.org/10.1088/0034-4885/75/7/076501} {\bibfield
   {journal} {\bibinfo  {journal} {Reports on Progress in Physics}\ }\textbf
  {\bibinfo {volume} {75}},\ \bibinfo {pages} {076501} (\bibinfo {year}
  {2012})}\BibitemShut {NoStop}%
\bibitem [{\citenamefont {Kawakami}\ and\ \citenamefont
  {Hu}(2015)}]{PhysRevLett.115.177001}%
  \BibitemOpen
  \bibfield  {author} {\bibinfo {author} {\bibfnamefont {T.}~\bibnamefont
  {Kawakami}}\ and\ \bibinfo {author} {\bibfnamefont {X.}~\bibnamefont {Hu}},\
  }\href {https://doi.org/10.1103/PhysRevLett.115.177001} {\bibfield  {journal}
  {\bibinfo  {journal} {Phys. Rev. Lett.}\ }\textbf {\bibinfo {volume} {115}},\
  \bibinfo {pages} {177001} (\bibinfo {year} {2015})}\BibitemShut {NoStop}%
\bibitem [{\citenamefont {Nadj-Perge}\ \emph {et~al.}(2014)\citenamefont
  {Nadj-Perge}, \citenamefont {Drozdov}, \citenamefont {Li}, \citenamefont
  {Chen}, \citenamefont {Jeon}, \citenamefont {Seo}, \citenamefont {MacDonald},
  \citenamefont {Bernevig},\ and\ \citenamefont
  {Yazdani}}]{doi:10.1126/science.1259327}%
  \BibitemOpen
  \bibfield  {author} {\bibinfo {author} {\bibfnamefont {S.}~\bibnamefont
  {Nadj-Perge}}, \bibinfo {author} {\bibfnamefont {I.~K.}\ \bibnamefont
  {Drozdov}}, \bibinfo {author} {\bibfnamefont {J.}~\bibnamefont {Li}},
  \bibinfo {author} {\bibfnamefont {H.}~\bibnamefont {Chen}}, \bibinfo {author}
  {\bibfnamefont {S.}~\bibnamefont {Jeon}}, \bibinfo {author} {\bibfnamefont
  {J.}~\bibnamefont {Seo}}, \bibinfo {author} {\bibfnamefont {A.~H.}\
  \bibnamefont {MacDonald}}, \bibinfo {author} {\bibfnamefont {B.~A.}\
  \bibnamefont {Bernevig}},\ and\ \bibinfo {author} {\bibfnamefont
  {A.}~\bibnamefont {Yazdani}},\ }\href
  {https://doi.org/10.1126/science.1259327} {\bibfield  {journal} {\bibinfo
  {journal} {Science}\ }\textbf {\bibinfo {volume} {346}},\ \bibinfo {pages}
  {602} (\bibinfo {year} {2014})}\BibitemShut {NoStop}%
\bibitem [{\citenamefont {Stone}\ and\ \citenamefont
  {Chung}(2006)}]{PhysRevB.73.014505}%
  \BibitemOpen
  \bibfield  {author} {\bibinfo {author} {\bibfnamefont {M.}~\bibnamefont
  {Stone}}\ and\ \bibinfo {author} {\bibfnamefont {S.-B.}\ \bibnamefont
  {Chung}},\ }\href {https://doi.org/10.1103/PhysRevB.73.014505} {\bibfield
  {journal} {\bibinfo  {journal} {Phys. Rev. B}\ }\textbf {\bibinfo {volume}
  {73}},\ \bibinfo {pages} {014505} (\bibinfo {year} {2006})}\BibitemShut
  {NoStop}%
\bibitem [{\citenamefont {Souto}\ and\ \citenamefont
  {Leijnse}(2022)}]{10.21468/SciPostPhys.12.5.161}%
  \BibitemOpen
  \bibfield  {author} {\bibinfo {author} {\bibfnamefont {R.~S.}\ \bibnamefont
  {Souto}}\ and\ \bibinfo {author} {\bibfnamefont {M.}~\bibnamefont
  {Leijnse}},\ }\href {https://doi.org/10.21468/SciPostPhys.12.5.161}
  {\bibfield  {journal} {\bibinfo  {journal} {SciPost Phys.}\ }\textbf
  {\bibinfo {volume} {12}},\ \bibinfo {pages} {161} (\bibinfo {year}
  {2022})}\BibitemShut {NoStop}%
\bibitem [{\citenamefont {Zhou}\ \emph {et~al.}(2022)\citenamefont {Zhou},
  \citenamefont {Dartiailh}, \citenamefont {Sardashti}, \citenamefont {Han},
  \citenamefont {Matos-Abiague}, \citenamefont {Shabani},\ and\ \citenamefont
  {{\v{Z}}uti{\'c}}}]{zhou2022fusion}%
  \BibitemOpen
  \bibfield  {author} {\bibinfo {author} {\bibfnamefont {T.}~\bibnamefont
  {Zhou}}, \bibinfo {author} {\bibfnamefont {M.~C.}\ \bibnamefont {Dartiailh}},
  \bibinfo {author} {\bibfnamefont {K.}~\bibnamefont {Sardashti}}, \bibinfo
  {author} {\bibfnamefont {J.~E.}\ \bibnamefont {Han}}, \bibinfo {author}
  {\bibfnamefont {A.}~\bibnamefont {Matos-Abiague}}, \bibinfo {author}
  {\bibfnamefont {J.}~\bibnamefont {Shabani}},\ and\ \bibinfo {author}
  {\bibfnamefont {I.}~\bibnamefont {{\v{Z}}uti{\'c}}},\ }\href
  {https://www.nature.com/articles/s41467-022-29463-6} {\bibfield  {journal}
  {\bibinfo  {journal} {Nature Communications}\ }\textbf {\bibinfo {volume}
  {13}},\ \bibinfo {pages} {1738} (\bibinfo {year} {2022})}\BibitemShut
  {NoStop}%
\bibitem [{\citenamefont {Ptok}\ \emph {et~al.}(2017)\citenamefont {Ptok},
  \citenamefont {Kobia\l{}ka},\ and\ \citenamefont {Doma\ifmmode~\acute{n}\else
  \'{n}\fi{}ski}}]{PhysRevB.96.195430}%
  \BibitemOpen
  \bibfield  {author} {\bibinfo {author} {\bibfnamefont {A.}~\bibnamefont
  {Ptok}}, \bibinfo {author} {\bibfnamefont {A.}~\bibnamefont {Kobia\l{}ka}},\
  and\ \bibinfo {author} {\bibfnamefont {T.}~\bibnamefont
  {Doma\ifmmode~\acute{n}\else \'{n}\fi{}ski}},\ }\href
  {https://doi.org/10.1103/PhysRevB.96.195430} {\bibfield  {journal} {\bibinfo
  {journal} {Phys. Rev. B}\ }\textbf {\bibinfo {volume} {96}},\ \bibinfo
  {pages} {195430} (\bibinfo {year} {2017})}\BibitemShut {NoStop}%
\bibitem [{\citenamefont {Kitaev}(2001)}]{Kitaev_2001}%
  \BibitemOpen
  \bibfield  {author} {\bibinfo {author} {\bibfnamefont {A.~Y.}\ \bibnamefont
  {Kitaev}},\ }\href {https://doi.org/10.1070/1063-7869/44/10s/s29} {\bibfield
  {journal} {\bibinfo  {journal} {Physics-Uspekhi}\ }\textbf {\bibinfo {volume}
  {44}},\ \bibinfo {pages} {131} (\bibinfo {year} {2001})}\BibitemShut
  {NoStop}%
\bibitem [{\citenamefont {Ivanov}(2001)}]{PhysRevLett.86.268}%
  \BibitemOpen
  \bibfield  {author} {\bibinfo {author} {\bibfnamefont {D.~A.}\ \bibnamefont
  {Ivanov}},\ }\href {https://doi.org/10.1103/PhysRevLett.86.268} {\bibfield
  {journal} {\bibinfo  {journal} {Phys. Rev. Lett.}\ }\textbf {\bibinfo
  {volume} {86}},\ \bibinfo {pages} {268} (\bibinfo {year} {2001})}\BibitemShut
  {NoStop}%
\bibitem [{\citenamefont {Feng}\ \emph {et~al.}(2018)\citenamefont {Feng},
  \citenamefont {Huang}, \citenamefont {Wang},\ and\ \citenamefont
  {Niu}}]{PhysRevB.98.134515}%
  \BibitemOpen
  \bibfield  {author} {\bibinfo {author} {\bibfnamefont {J.-J.}\ \bibnamefont
  {Feng}}, \bibinfo {author} {\bibfnamefont {Z.}~\bibnamefont {Huang}},
  \bibinfo {author} {\bibfnamefont {Z.}~\bibnamefont {Wang}},\ and\ \bibinfo
  {author} {\bibfnamefont {Q.}~\bibnamefont {Niu}},\ }\href
  {https://doi.org/10.1103/PhysRevB.98.134515} {\bibfield  {journal} {\bibinfo
  {journal} {Phys. Rev. B}\ }\textbf {\bibinfo {volume} {98}},\ \bibinfo
  {pages} {134515} (\bibinfo {year} {2018})}\BibitemShut {NoStop}%
\bibitem [{\citenamefont {Feng}\ \emph {et~al.}(2020)\citenamefont {Feng},
  \citenamefont {Huang}, \citenamefont {Wang},\ and\ \citenamefont
  {Niu}}]{PhysRevB.101.180504}%
  \BibitemOpen
  \bibfield  {author} {\bibinfo {author} {\bibfnamefont {J.-J.}\ \bibnamefont
  {Feng}}, \bibinfo {author} {\bibfnamefont {Z.}~\bibnamefont {Huang}},
  \bibinfo {author} {\bibfnamefont {Z.}~\bibnamefont {Wang}},\ and\ \bibinfo
  {author} {\bibfnamefont {Q.}~\bibnamefont {Niu}},\ }\href
  {https://doi.org/10.1103/PhysRevB.101.180504} {\bibfield  {journal} {\bibinfo
   {journal} {Phys. Rev. B}\ }\textbf {\bibinfo {volume} {101}},\ \bibinfo
  {pages} {180504} (\bibinfo {year} {2020})}\BibitemShut {NoStop}%
\bibitem [{\citenamefont {Xu}\ \emph {et~al.}(2017)\citenamefont {Xu},
  \citenamefont {Li},\ and\ \citenamefont {Sun}}]{Xu_2017}%
  \BibitemOpen
  \bibfield  {author} {\bibinfo {author} {\bibfnamefont {L.}~\bibnamefont
  {Xu}}, \bibinfo {author} {\bibfnamefont {X.-Q.}\ \bibnamefont {Li}},\ and\
  \bibinfo {author} {\bibfnamefont {Q.-F.}\ \bibnamefont {Sun}},\ }\href
  {https://doi.org/10.1088/1361-648x/aa6661} {\bibfield  {journal} {\bibinfo
  {journal} {Journal of Physics: Condensed Matter}\ }\textbf {\bibinfo {volume}
  {29}},\ \bibinfo {pages} {195301} (\bibinfo {year} {2017})}\BibitemShut
  {NoStop}%
\bibitem [{\citenamefont {Choi}\ \emph {et~al.}(2020)\citenamefont {Choi},
  \citenamefont {Calzona},\ and\ \citenamefont
  {Trauzettel}}]{PhysRevB.102.140501}%
  \BibitemOpen
  \bibfield  {author} {\bibinfo {author} {\bibfnamefont {S.-J.}\ \bibnamefont
  {Choi}}, \bibinfo {author} {\bibfnamefont {A.}~\bibnamefont {Calzona}},\ and\
  \bibinfo {author} {\bibfnamefont {B.}~\bibnamefont {Trauzettel}},\ }\href
  {https://doi.org/10.1103/PhysRevB.102.140501} {\bibfield  {journal} {\bibinfo
   {journal} {Phys. Rev. B}\ }\textbf {\bibinfo {volume} {102}},\ \bibinfo
  {pages} {140501} (\bibinfo {year} {2020})}\BibitemShut {NoStop}%
\bibitem [{\citenamefont {Oriekhov}\ \emph {et~al.}(2021)\citenamefont
  {Oriekhov}, \citenamefont {Cheipesh},\ and\ \citenamefont
  {Beenakker}}]{PhysRevB.103.094518}%
  \BibitemOpen
  \bibfield  {author} {\bibinfo {author} {\bibfnamefont {D.~O.}\ \bibnamefont
  {Oriekhov}}, \bibinfo {author} {\bibfnamefont {Y.}~\bibnamefont {Cheipesh}},\
  and\ \bibinfo {author} {\bibfnamefont {C.~W.~J.}\ \bibnamefont {Beenakker}},\
  }\href {https://doi.org/10.1103/PhysRevB.103.094518} {\bibfield  {journal}
  {\bibinfo  {journal} {Phys. Rev. B}\ }\textbf {\bibinfo {volume} {103}},\
  \bibinfo {pages} {094518} (\bibinfo {year} {2021})}\BibitemShut {NoStop}%
\bibitem [{\citenamefont {Parafilo}\ and\ \citenamefont
  {Kiselev}(2018)}]{doi:10.1063/1.5078628}%
  \BibitemOpen
  \bibfield  {author} {\bibinfo {author} {\bibfnamefont {A.~V.}\ \bibnamefont
  {Parafilo}}\ and\ \bibinfo {author} {\bibfnamefont {M.~N.}\ \bibnamefont
  {Kiselev}},\ }\href {https://doi.org/10.1063/1.5078628} {\bibfield  {journal}
  {\bibinfo  {journal} {Low Temperature Physics}\ }\textbf {\bibinfo {volume}
  {44}},\ \bibinfo {pages} {1325} (\bibinfo {year} {2018})}\BibitemShut
  {NoStop}%
\bibitem [{\citenamefont {Schön}\ \emph {et~al.}(2020)\citenamefont {Schön},
  \citenamefont {Voss}, \citenamefont {Wildermuth}, \citenamefont {Schneider},
  \citenamefont {Skacel}, \citenamefont {Weides}, \citenamefont {Cole},
  \citenamefont {Rotzinger},\ and\ \citenamefont {Ustinov}}]{Schon2020}%
  \BibitemOpen
  \bibfield  {author} {\bibinfo {author} {\bibfnamefont {Y.}~\bibnamefont
  {Schön}}, \bibinfo {author} {\bibfnamefont {J.~N.}\ \bibnamefont {Voss}},
  \bibinfo {author} {\bibfnamefont {M.}~\bibnamefont {Wildermuth}}, \bibinfo
  {author} {\bibfnamefont {A.}~\bibnamefont {Schneider}}, \bibinfo {author}
  {\bibfnamefont {S.~T.}\ \bibnamefont {Skacel}}, \bibinfo {author}
  {\bibfnamefont {M.~P.}\ \bibnamefont {Weides}}, \bibinfo {author}
  {\bibfnamefont {J.~H.}\ \bibnamefont {Cole}}, \bibinfo {author}
  {\bibfnamefont {H.}~\bibnamefont {Rotzinger}},\ and\ \bibinfo {author}
  {\bibfnamefont {A.~V.}\ \bibnamefont {Ustinov}},\ }\href
  {https://doi.org/https://doi.org/10.1038/s41535-020-0220-x} {\bibfield
  {journal} {\bibinfo  {journal} {npj Quantum Mater}\ }\textbf {\bibinfo
  {volume} {5}},\ \bibinfo {pages} {18} (\bibinfo {year} {2020})}\BibitemShut
  {NoStop}%
\bibitem [{\citenamefont {Shevchenko}\ \emph {et~al.}(2008)\citenamefont
  {Shevchenko}, \citenamefont {Omelyanchouk}, \citenamefont {Zagoskin},
  \citenamefont {Savel{\textquotesingle}ev},\ and\ \citenamefont
  {Nori}}]{Shevchenko_2008}%
  \BibitemOpen
  \bibfield  {author} {\bibinfo {author} {\bibfnamefont {S.~N.}\ \bibnamefont
  {Shevchenko}}, \bibinfo {author} {\bibfnamefont {A.~N.}\ \bibnamefont
  {Omelyanchouk}}, \bibinfo {author} {\bibfnamefont {A.~M.}\ \bibnamefont
  {Zagoskin}}, \bibinfo {author} {\bibfnamefont {S.}~\bibnamefont
  {Savel{\textquotesingle}ev}},\ and\ \bibinfo {author} {\bibfnamefont
  {F.}~\bibnamefont {Nori}},\ }\href
  {https://doi.org/10.1088/1367-2630/10/7/073026} {\bibfield  {journal}
  {\bibinfo  {journal} {New Journal of Physics}\ }\textbf {\bibinfo {volume}
  {10}},\ \bibinfo {pages} {073026} (\bibinfo {year} {2008})}\BibitemShut
  {NoStop}%
\bibitem [{\citenamefont {Gr\o{}nbech-Jensen}\ and\ \citenamefont
  {Cirillo}(2005)}]{PhysRevLett.95.067001}%
  \BibitemOpen
  \bibfield  {author} {\bibinfo {author} {\bibfnamefont {N.}~\bibnamefont
  {Gr\o{}nbech-Jensen}}\ and\ \bibinfo {author} {\bibfnamefont
  {M.}~\bibnamefont {Cirillo}},\ }\href
  {https://doi.org/10.1103/PhysRevLett.95.067001} {\bibfield  {journal}
  {\bibinfo  {journal} {Phys. Rev. Lett.}\ }\textbf {\bibinfo {volume} {95}},\
  \bibinfo {pages} {067001} (\bibinfo {year} {2005})}\BibitemShut {NoStop}%
\bibitem [{\citenamefont {Dutta}\ \emph {et~al.}(2008)\citenamefont {Dutta},
  \citenamefont {Strauch}, \citenamefont {Lewis}, \citenamefont {Mitra},
  \citenamefont {Paik}, \citenamefont {Palomaki}, \citenamefont {Tiesinga},
  \citenamefont {Anderson}, \citenamefont {Dragt}, \citenamefont {Lobb},\ and\
  \citenamefont {Wellstood}}]{PhysRevB.78.104510}%
  \BibitemOpen
  \bibfield  {author} {\bibinfo {author} {\bibfnamefont {S.~K.}\ \bibnamefont
  {Dutta}}, \bibinfo {author} {\bibfnamefont {F.~W.}\ \bibnamefont {Strauch}},
  \bibinfo {author} {\bibfnamefont {R.~M.}\ \bibnamefont {Lewis}}, \bibinfo
  {author} {\bibfnamefont {K.}~\bibnamefont {Mitra}}, \bibinfo {author}
  {\bibfnamefont {H.}~\bibnamefont {Paik}}, \bibinfo {author} {\bibfnamefont
  {T.~A.}\ \bibnamefont {Palomaki}}, \bibinfo {author} {\bibfnamefont
  {E.}~\bibnamefont {Tiesinga}}, \bibinfo {author} {\bibfnamefont {J.~R.}\
  \bibnamefont {Anderson}}, \bibinfo {author} {\bibfnamefont {A.~J.}\
  \bibnamefont {Dragt}}, \bibinfo {author} {\bibfnamefont {C.~J.}\ \bibnamefont
  {Lobb}},\ and\ \bibinfo {author} {\bibfnamefont {F.~C.}\ \bibnamefont
  {Wellstood}},\ }\href {https://doi.org/10.1103/PhysRevB.78.104510} {\bibfield
   {journal} {\bibinfo  {journal} {Phys. Rev. B}\ }\textbf {\bibinfo {volume}
  {78}},\ \bibinfo {pages} {104510} (\bibinfo {year} {2008})}\BibitemShut
  {NoStop}%
\bibitem [{\citenamefont {Majorana}(1937)}]{majorana1937teoria}%
  \BibitemOpen
  \bibfield  {author} {\bibinfo {author} {\bibfnamefont {E.}~\bibnamefont
  {Majorana}},\ }\href {https://doi.org/10.1007/BF02961314} {\bibfield
  {journal} {\bibinfo  {journal} {Il Nuovo Cimento (1924-1942)}\ }\textbf
  {\bibinfo {volume} {14}},\ \bibinfo {pages} {171} (\bibinfo {year}
  {1937})}\BibitemShut {NoStop}%
\bibitem [{\citenamefont {Kwon}\ \emph {et~al.}(2004)\citenamefont {Kwon},
  \citenamefont {Sengupta},\ and\ \citenamefont
  {Yakovenko}}]{kwon2004fractional}%
  \BibitemOpen
  \bibfield  {author} {\bibinfo {author} {\bibfnamefont {H.-J.}\ \bibnamefont
  {Kwon}}, \bibinfo {author} {\bibfnamefont {K.}~\bibnamefont {Sengupta}},\
  and\ \bibinfo {author} {\bibfnamefont {V.~M.}\ \bibnamefont {Yakovenko}},\
  }\href {https://doi.org/10.1140/epjb/e2004-00066-4} {\bibfield  {journal}
  {\bibinfo  {journal} {The European Physical Journal B-Condensed Matter and
  Complex Systems}\ }\textbf {\bibinfo {volume} {37}},\ \bibinfo {pages} {349}
  (\bibinfo {year} {2004})}\BibitemShut {NoStop}%
\bibitem [{\citenamefont {Rokhinson}\ \emph {et~al.}(2012)\citenamefont
  {Rokhinson}, \citenamefont {Liu},\ and\ \citenamefont
  {Furdyna}}]{rokhinson2012fractional}%
  \BibitemOpen
  \bibfield  {author} {\bibinfo {author} {\bibfnamefont {L.~P.}\ \bibnamefont
  {Rokhinson}}, \bibinfo {author} {\bibfnamefont {X.}~\bibnamefont {Liu}},\
  and\ \bibinfo {author} {\bibfnamefont {J.~K.}\ \bibnamefont {Furdyna}},\
  }\href {https://doi.org/10.1038/nphys2429} {\bibfield  {journal} {\bibinfo
  {journal} {Nature Physics}\ }\textbf {\bibinfo {volume} {8}},\ \bibinfo
  {pages} {795} (\bibinfo {year} {2012})}\BibitemShut {NoStop}%
\bibitem [{\citenamefont {Deacon}\ \emph {et~al.}(2017)\citenamefont {Deacon},
  \citenamefont {Wiedenmann}, \citenamefont {Bocquillon}, \citenamefont
  {Dom\'{\i}nguez}, \citenamefont {Klapwijk}, \citenamefont {Leubner},
  \citenamefont {Br\"une}, \citenamefont {Hankiewicz}, \citenamefont {Tarucha},
  \citenamefont {Ishibashi}, \citenamefont {Buhmann},\ and\ \citenamefont
  {Molenkamp}}]{PhysRevX.7.021011}%
  \BibitemOpen
  \bibfield  {author} {\bibinfo {author} {\bibfnamefont {R.~S.}\ \bibnamefont
  {Deacon}}, \bibinfo {author} {\bibfnamefont {J.}~\bibnamefont {Wiedenmann}},
  \bibinfo {author} {\bibfnamefont {E.}~\bibnamefont {Bocquillon}}, \bibinfo
  {author} {\bibfnamefont {F.}~\bibnamefont {Dom\'{\i}nguez}}, \bibinfo
  {author} {\bibfnamefont {T.~M.}\ \bibnamefont {Klapwijk}}, \bibinfo {author}
  {\bibfnamefont {P.}~\bibnamefont {Leubner}}, \bibinfo {author} {\bibfnamefont
  {C.}~\bibnamefont {Br\"une}}, \bibinfo {author} {\bibfnamefont {E.~M.}\
  \bibnamefont {Hankiewicz}}, \bibinfo {author} {\bibfnamefont
  {S.}~\bibnamefont {Tarucha}}, \bibinfo {author} {\bibfnamefont
  {K.}~\bibnamefont {Ishibashi}}, \bibinfo {author} {\bibfnamefont
  {H.}~\bibnamefont {Buhmann}},\ and\ \bibinfo {author} {\bibfnamefont {L.~W.}\
  \bibnamefont {Molenkamp}},\ }\href
  {https://doi.org/10.1103/PhysRevX.7.021011} {\bibfield  {journal} {\bibinfo
  {journal} {Phys. Rev. X}\ }\textbf {\bibinfo {volume} {7}},\ \bibinfo {pages}
  {021011} (\bibinfo {year} {2017})}\BibitemShut {NoStop}%
\bibitem [{\citenamefont {Landau}(1932)}]{landau1932b}%
  \BibitemOpen
  \bibfield  {author} {\bibinfo {author} {\bibfnamefont {L.~D.}\ \bibnamefont
  {Landau}},\ }\href@noop {} {\bibfield  {journal} {\bibinfo  {journal} {Phys.
  Z. Sowjetunion}\ }\textbf {\bibinfo {volume} {2}},\ \bibinfo {pages} {46}
  (\bibinfo {year} {1932})}\BibitemShut {NoStop}%
\bibitem [{\citenamefont {Zener}(1932)}]{zener1932}%
  \BibitemOpen
  \bibfield  {author} {\bibinfo {author} {\bibfnamefont {C.}~\bibnamefont
  {Zener}},\ }\href {https://doi.org/10.1098/rspa.1932.0165} {\bibfield
  {journal} {\bibinfo  {journal} {Proc. R. Soc. London, Ser. A}\ }\textbf
  {\bibinfo {volume} {137}},\ \bibinfo {pages} {696} (\bibinfo {year}
  {1932})}\BibitemShut {NoStop}%
\bibitem [{\citenamefont {Peng}\ \emph {et~al.}(2015)\citenamefont {Peng},
  \citenamefont {Pientka}, \citenamefont {Vinkler-Aviv}, \citenamefont
  {Glazman},\ and\ \citenamefont {von Oppen}}]{PhysRevLett.115.266804}%
  \BibitemOpen
  \bibfield  {author} {\bibinfo {author} {\bibfnamefont {Y.}~\bibnamefont
  {Peng}}, \bibinfo {author} {\bibfnamefont {F.}~\bibnamefont {Pientka}},
  \bibinfo {author} {\bibfnamefont {Y.}~\bibnamefont {Vinkler-Aviv}}, \bibinfo
  {author} {\bibfnamefont {L.~I.}\ \bibnamefont {Glazman}},\ and\ \bibinfo
  {author} {\bibfnamefont {F.}~\bibnamefont {von Oppen}},\ }\href
  {https://doi.org/10.1103/PhysRevLett.115.266804} {\bibfield  {journal}
  {\bibinfo  {journal} {Phys. Rev. Lett.}\ }\textbf {\bibinfo {volume} {115}},\
  \bibinfo {pages} {266804} (\bibinfo {year} {2015})}\BibitemShut {NoStop}%
\bibitem [{\citenamefont {Kobiałka}\ and\ \citenamefont
  {Ptok}(2020)}]{Kobialka2020681}%
  \BibitemOpen
  \bibfield  {author} {\bibinfo {author} {\bibfnamefont {A.}~\bibnamefont
  {Kobiałka}}\ and\ \bibinfo {author} {\bibfnamefont {A.}~\bibnamefont
  {Ptok}},\ }\href {https://doi.org/10.12693/APHYSPOLA.138.681} {\bibfield
  {journal} {\bibinfo  {journal} {Acta Physica Polonica A}\ }\textbf {\bibinfo
  {volume} {138}},\ \bibinfo {pages} {681} (\bibinfo {year}
  {2020})}\BibitemShut {NoStop}%
\bibitem [{\citenamefont {Liu}\ \emph {et~al.}(2020)\citenamefont {Liu},
  \citenamefont {Chen}, \citenamefont {Song},\ and\ \citenamefont
  {Li}}]{LIU2020114212}%
  \BibitemOpen
  \bibfield  {author} {\bibinfo {author} {\bibfnamefont {L.}~\bibnamefont
  {Liu}}, \bibinfo {author} {\bibfnamefont {X.-F.}\ \bibnamefont {Chen}},
  \bibinfo {author} {\bibfnamefont {J.}~\bibnamefont {Song}},\ and\ \bibinfo
  {author} {\bibfnamefont {Y.-X.}\ \bibnamefont {Li}},\ }\href
  {https://doi.org/https://doi.org/10.1016/j.physe.2020.114212} {\bibfield
  {journal} {\bibinfo  {journal} {Physica E: Low-dimensional Systems and
  Nanostructures}\ }\textbf {\bibinfo {volume} {124}},\ \bibinfo {pages}
  {114212} (\bibinfo {year} {2020})}\BibitemShut {NoStop}%
\bibitem [{\citenamefont {Yao}\ and\ \citenamefont {Zhang}(2022)}]{yao_2022}%
  \BibitemOpen
  \bibfield  {author} {\bibinfo {author} {\bibfnamefont {C.-Z.}\ \bibnamefont
  {Yao}}\ and\ \bibinfo {author} {\bibfnamefont {W.-M.}\ \bibnamefont
  {Zhang}},\ }\href {https://doi.org/10.1088/1367-2630/ac7c85} {\bibfield
  {journal} {\bibinfo  {journal} {New Journal of Physics}\ }\textbf {\bibinfo
  {volume} {24}},\ \bibinfo {pages} {073015} (\bibinfo {year}
  {2022})}\BibitemShut {NoStop}%
\bibitem [{\citenamefont {Fu}\ and\ \citenamefont
  {Kane}(2009)}]{PhysRevB.79.161408}%
  \BibitemOpen
  \bibfield  {author} {\bibinfo {author} {\bibfnamefont {L.}~\bibnamefont
  {Fu}}\ and\ \bibinfo {author} {\bibfnamefont {C.~L.}\ \bibnamefont {Kane}},\
  }\href {https://doi.org/10.1103/PhysRevB.79.161408} {\bibfield  {journal}
  {\bibinfo  {journal} {Phys. Rev. B}\ }\textbf {\bibinfo {volume} {79}},\
  \bibinfo {pages} {161408} (\bibinfo {year} {2009})}\BibitemShut {NoStop}%
\bibitem [{\citenamefont {Peng}\ \emph {et~al.}(2016)\citenamefont {Peng},
  \citenamefont {Pientka}, \citenamefont {Berg}, \citenamefont {Oreg},\ and\
  \citenamefont {von Oppen}}]{PhysRevB.94.085409}%
  \BibitemOpen
  \bibfield  {author} {\bibinfo {author} {\bibfnamefont {Y.}~\bibnamefont
  {Peng}}, \bibinfo {author} {\bibfnamefont {F.}~\bibnamefont {Pientka}},
  \bibinfo {author} {\bibfnamefont {E.}~\bibnamefont {Berg}}, \bibinfo {author}
  {\bibfnamefont {Y.}~\bibnamefont {Oreg}},\ and\ \bibinfo {author}
  {\bibfnamefont {F.}~\bibnamefont {von Oppen}},\ }\href
  {https://doi.org/10.1103/PhysRevB.94.085409} {\bibfield  {journal} {\bibinfo
  {journal} {Phys. Rev. B}\ }\textbf {\bibinfo {volume} {94}},\ \bibinfo
  {pages} {085409} (\bibinfo {year} {2016})}\BibitemShut {NoStop}%
\bibitem [{\citenamefont {Rainis}\ and\ \citenamefont
  {Loss}(2012)}]{PhysRevB.85.174533}%
  \BibitemOpen
  \bibfield  {author} {\bibinfo {author} {\bibfnamefont {D.}~\bibnamefont
  {Rainis}}\ and\ \bibinfo {author} {\bibfnamefont {D.}~\bibnamefont {Loss}},\
  }\href {https://doi.org/10.1103/PhysRevB.85.174533} {\bibfield  {journal}
  {\bibinfo  {journal} {Phys. Rev. B}\ }\textbf {\bibinfo {volume} {85}},\
  \bibinfo {pages} {174533} (\bibinfo {year} {2012})}\BibitemShut {NoStop}%
\bibitem [{\citenamefont {Schmidt}\ \emph {et~al.}(2012)\citenamefont
  {Schmidt}, \citenamefont {Rainis},\ and\ \citenamefont
  {Loss}}]{PhysRevB.86.085414}%
  \BibitemOpen
  \bibfield  {author} {\bibinfo {author} {\bibfnamefont {M.~J.}\ \bibnamefont
  {Schmidt}}, \bibinfo {author} {\bibfnamefont {D.}~\bibnamefont {Rainis}},\
  and\ \bibinfo {author} {\bibfnamefont {D.}~\bibnamefont {Loss}},\ }\href
  {https://doi.org/10.1103/PhysRevB.86.085414} {\bibfield  {journal} {\bibinfo
  {journal} {Phys. Rev. B}\ }\textbf {\bibinfo {volume} {86}},\ \bibinfo
  {pages} {085414} (\bibinfo {year} {2012})}\BibitemShut {NoStop}%
\bibitem [{\citenamefont {Colbert}\ and\ \citenamefont
  {Lee}(2014)}]{PhysRevB.89.140505}%
  \BibitemOpen
  \bibfield  {author} {\bibinfo {author} {\bibfnamefont {J.~R.}\ \bibnamefont
  {Colbert}}\ and\ \bibinfo {author} {\bibfnamefont {P.~A.}\ \bibnamefont
  {Lee}},\ }\href {https://doi.org/10.1103/PhysRevB.89.140505} {\bibfield
  {journal} {\bibinfo  {journal} {Phys. Rev. B}\ }\textbf {\bibinfo {volume}
  {89}},\ \bibinfo {pages} {140505} (\bibinfo {year} {2014})}\BibitemShut
  {NoStop}%
\bibitem [{\citenamefont {Albrecht}\ \emph {et~al.}(2017)\citenamefont
  {Albrecht}, \citenamefont {Hansen}, \citenamefont {Higginbotham},
  \citenamefont {Kuemmeth}, \citenamefont {Jespersen}, \citenamefont
  {Nyg\aa{}rd}, \citenamefont {Krogstrup}, \citenamefont {Danon}, \citenamefont
  {Flensberg},\ and\ \citenamefont {Marcus}}]{PhysRevLett.118.137701}%
  \BibitemOpen
  \bibfield  {author} {\bibinfo {author} {\bibfnamefont {S.~M.}\ \bibnamefont
  {Albrecht}}, \bibinfo {author} {\bibfnamefont {E.~B.}\ \bibnamefont
  {Hansen}}, \bibinfo {author} {\bibfnamefont {A.~P.}\ \bibnamefont
  {Higginbotham}}, \bibinfo {author} {\bibfnamefont {F.}~\bibnamefont
  {Kuemmeth}}, \bibinfo {author} {\bibfnamefont {T.~S.}\ \bibnamefont
  {Jespersen}}, \bibinfo {author} {\bibfnamefont {J.}~\bibnamefont
  {Nyg\aa{}rd}}, \bibinfo {author} {\bibfnamefont {P.}~\bibnamefont
  {Krogstrup}}, \bibinfo {author} {\bibfnamefont {J.}~\bibnamefont {Danon}},
  \bibinfo {author} {\bibfnamefont {K.}~\bibnamefont {Flensberg}},\ and\
  \bibinfo {author} {\bibfnamefont {C.~M.}\ \bibnamefont {Marcus}},\ }\href
  {https://doi.org/10.1103/PhysRevLett.118.137701} {\bibfield  {journal}
  {\bibinfo  {journal} {Phys. Rev. Lett.}\ }\textbf {\bibinfo {volume} {118}},\
  \bibinfo {pages} {137701} (\bibinfo {year} {2017})}\BibitemShut {NoStop}%
\bibitem [{\citenamefont {Xiang}\ \emph {et~al.}(2013)\citenamefont {Xiang},
  \citenamefont {Ashhab}, \citenamefont {You},\ and\ \citenamefont
  {Nori}}]{RevModPhys.85.623}%
  \BibitemOpen
  \bibfield  {author} {\bibinfo {author} {\bibfnamefont {Z.-L.}\ \bibnamefont
  {Xiang}}, \bibinfo {author} {\bibfnamefont {S.}~\bibnamefont {Ashhab}},
  \bibinfo {author} {\bibfnamefont {J.~Q.}\ \bibnamefont {You}},\ and\ \bibinfo
  {author} {\bibfnamefont {F.}~\bibnamefont {Nori}},\ }\href
  {https://doi.org/10.1103/RevModPhys.85.623} {\bibfield  {journal} {\bibinfo
  {journal} {Rev. Mod. Phys.}\ }\textbf {\bibinfo {volume} {85}},\ \bibinfo
  {pages} {623} (\bibinfo {year} {2013})}\BibitemShut {NoStop}%
\bibitem [{\citenamefont {Devoret}\ \emph {et~al.}(2004)\citenamefont
  {Devoret}, \citenamefont {Wallraff},\ and\ \citenamefont
  {Martinis}}]{https://doi.org/10.48550/arxiv.cond-mat/0411174}%
  \BibitemOpen
  \bibfield  {author} {\bibinfo {author} {\bibfnamefont {M.~H.}\ \bibnamefont
  {Devoret}}, \bibinfo {author} {\bibfnamefont {A.}~\bibnamefont {Wallraff}},\
  and\ \bibinfo {author} {\bibfnamefont {J.~M.}\ \bibnamefont {Martinis}},\
  }\href {https://doi.org/10.48550/ARXIV.COND-MAT/0411174} {\bibinfo {title}
  {Superconducting qubits: A short review}} (\bibinfo {year}
  {2004})\BibitemShut {NoStop}%
\bibitem [{\citenamefont {Breuer}\ and\ \citenamefont
  {Petruccione}(2002)}]{Breuer2002TheTheoryOfOpenQuantumSystems}%
  \BibitemOpen
  \bibfield  {author} {\bibinfo {author} {\bibfnamefont {H.-P.}\ \bibnamefont
  {Breuer}}\ and\ \bibinfo {author} {\bibfnamefont {F.}~\bibnamefont
  {Petruccione}},\ }\href {https://academic.oup.com/book/27757} {\emph
  {\bibinfo {title} {The Theory of Open Quantum Systems}}}\ (\bibinfo
  {publisher} {Oxford University Press},\ \bibinfo {year} {2002})\ pp.\
  \bibinfo {pages} {110--125}\BibitemShut {NoStop}%
\bibitem [{\citenamefont {Nielsen}\ and\ \citenamefont
  {Chuang}(2010)}]{nielsen_chuang_2010}%
  \BibitemOpen
  \bibfield  {author} {\bibinfo {author} {\bibfnamefont {M.~A.}\ \bibnamefont
  {Nielsen}}\ and\ \bibinfo {author} {\bibfnamefont {I.~L.}\ \bibnamefont
  {Chuang}},\ }\href {https://doi.org/10.1017/CBO9780511976667} {\emph
  {\bibinfo {title} {Quantum Computation and Quantum Information: 10th
  Anniversary Edition}}}\ (\bibinfo  {publisher} {Cambridge University Press},\
  \bibinfo {year} {2010})\ pp.\ \bibinfo {pages} {386--389}\BibitemShut
  {NoStop}%
\bibitem [{\citenamefont {Huang}\ \emph {et~al.}(2015)\citenamefont {Huang},
  \citenamefont {Liang}, \citenamefont {Yao},\ and\ \citenamefont
  {Wang}}]{PhysRevA.92.012308}%
  \BibitemOpen
  \bibfield  {author} {\bibinfo {author} {\bibfnamefont {W.-C.}\ \bibnamefont
  {Huang}}, \bibinfo {author} {\bibfnamefont {Q.-F.}\ \bibnamefont {Liang}},
  \bibinfo {author} {\bibfnamefont {D.-X.}\ \bibnamefont {Yao}},\ and\ \bibinfo
  {author} {\bibfnamefont {Z.}~\bibnamefont {Wang}},\ }\href
  {https://doi.org/10.1103/PhysRevA.92.012308} {\bibfield  {journal} {\bibinfo
  {journal} {Phys. Rev. A}\ }\textbf {\bibinfo {volume} {92}},\ \bibinfo
  {pages} {012308} (\bibinfo {year} {2015})}\BibitemShut {NoStop}%
\bibitem [{\citenamefont {Huang}\ \emph {et~al.}(2011)\citenamefont {Huang},
  \citenamefont {Kong}, \citenamefont {Zhao}, \citenamefont {Shi},
  \citenamefont {Wang}, \citenamefont {Rong}, \citenamefont {Liu},\ and\
  \citenamefont {Du}}]{huang2011observation}%
  \BibitemOpen
  \bibfield  {author} {\bibinfo {author} {\bibfnamefont {P.}~\bibnamefont
  {Huang}}, \bibinfo {author} {\bibfnamefont {X.}~\bibnamefont {Kong}},
  \bibinfo {author} {\bibfnamefont {N.}~\bibnamefont {Zhao}}, \bibinfo {author}
  {\bibfnamefont {F.}~\bibnamefont {Shi}}, \bibinfo {author} {\bibfnamefont
  {P.}~\bibnamefont {Wang}}, \bibinfo {author} {\bibfnamefont {X.}~\bibnamefont
  {Rong}}, \bibinfo {author} {\bibfnamefont {R.-B.}\ \bibnamefont {Liu}},\ and\
  \bibinfo {author} {\bibfnamefont {J.}~\bibnamefont {Du}},\ }\href
  {https://doi.org/10.1038/ncomms1579} {\bibfield  {journal} {\bibinfo
  {journal} {Nature communications}\ }\textbf {\bibinfo {volume} {2}},\
  \bibinfo {pages} {1} (\bibinfo {year} {2011})}\BibitemShut {NoStop}%
\bibitem [{\citenamefont {Hong}\ \emph {et~al.}(2013)\citenamefont {Hong},
  \citenamefont {Grinolds}, \citenamefont {Pham}, \citenamefont {Le~Sage},
  \citenamefont {Luan}, \citenamefont {Walsworth},\ and\ \citenamefont
  {Yacoby}}]{hong2013nanoscale}%
  \BibitemOpen
  \bibfield  {author} {\bibinfo {author} {\bibfnamefont {S.}~\bibnamefont
  {Hong}}, \bibinfo {author} {\bibfnamefont {M.~S.}\ \bibnamefont {Grinolds}},
  \bibinfo {author} {\bibfnamefont {L.~M.}\ \bibnamefont {Pham}}, \bibinfo
  {author} {\bibfnamefont {D.}~\bibnamefont {Le~Sage}}, \bibinfo {author}
  {\bibfnamefont {L.}~\bibnamefont {Luan}}, \bibinfo {author} {\bibfnamefont
  {R.~L.}\ \bibnamefont {Walsworth}},\ and\ \bibinfo {author} {\bibfnamefont
  {A.}~\bibnamefont {Yacoby}},\ }\href {https://doi.org/10.1557/mrs.2013.23}
  {\bibfield  {journal} {\bibinfo  {journal} {MRS bulletin}\ }\textbf {\bibinfo
  {volume} {38}},\ \bibinfo {pages} {155} (\bibinfo {year} {2013})}\BibitemShut
  {NoStop}%
\bibitem [{\citenamefont {Wang}\ \emph {et~al.}(2016)\citenamefont {Wang},
  \citenamefont {Zu}, \citenamefont {He}, \citenamefont {Wang}, \citenamefont
  {Zhang},\ and\ \citenamefont {Duan}}]{PhysRevB.94.064304}%
  \BibitemOpen
  \bibfield  {author} {\bibinfo {author} {\bibfnamefont {F.}~\bibnamefont
  {Wang}}, \bibinfo {author} {\bibfnamefont {C.}~\bibnamefont {Zu}}, \bibinfo
  {author} {\bibfnamefont {L.}~\bibnamefont {He}}, \bibinfo {author}
  {\bibfnamefont {W.-B.}\ \bibnamefont {Wang}}, \bibinfo {author}
  {\bibfnamefont {W.-G.}\ \bibnamefont {Zhang}},\ and\ \bibinfo {author}
  {\bibfnamefont {L.-M.}\ \bibnamefont {Duan}},\ }\href
  {https://doi.org/10.1103/PhysRevB.94.064304} {\bibfield  {journal} {\bibinfo
  {journal} {Phys. Rev. B}\ }\textbf {\bibinfo {volume} {94}},\ \bibinfo
  {pages} {064304} (\bibinfo {year} {2016})}\BibitemShut {NoStop}%
\bibitem [{\citenamefont {Shevchenko}\ \emph {et~al.}(2010)\citenamefont
  {Shevchenko}, \citenamefont {Ashhab},\ and\ \citenamefont
  {Nori}}]{SHEVCHENKO20101}%
  \BibitemOpen
  \bibfield  {author} {\bibinfo {author} {\bibfnamefont {S.}~\bibnamefont
  {Shevchenko}}, \bibinfo {author} {\bibfnamefont {S.}~\bibnamefont {Ashhab}},\
  and\ \bibinfo {author} {\bibfnamefont {F.}~\bibnamefont {Nori}},\ }\href
  {https://doi.org/https://doi.org/10.1016/j.physrep.2010.03.002} {\bibfield
  {journal} {\bibinfo  {journal} {Physics Reports}\ }\textbf {\bibinfo {volume}
  {492}},\ \bibinfo {pages} {1} (\bibinfo {year} {2010})}\BibitemShut {NoStop}%
\bibitem [{\citenamefont {Ivakhnenko}\ \emph {et~al.}(2023)\citenamefont
  {Ivakhnenko}, \citenamefont {Shevchenko},\ and\ \citenamefont
  {Nori}}]{IVAKHNENKO20231}%
  \BibitemOpen
  \bibfield  {author} {\bibinfo {author} {\bibfnamefont {O.~V.}\ \bibnamefont
  {Ivakhnenko}}, \bibinfo {author} {\bibfnamefont {S.~N.}\ \bibnamefont
  {Shevchenko}},\ and\ \bibinfo {author} {\bibfnamefont {F.}~\bibnamefont
  {Nori}},\ }\href
  {https://doi.org/https://doi.org/10.1016/j.physrep.2022.10.002} {\bibfield
  {journal} {\bibinfo  {journal} {Physics Reports}\ }\textbf {\bibinfo {volume}
  {995}},\ \bibinfo {pages} {1} (\bibinfo {year} {2023})}\BibitemShut {NoStop}%
\bibitem [{\citenamefont {Peng}\ and\ \citenamefont
  {Yang}(2013)}]{10.1360/132012-930}%
  \BibitemOpen
  \bibfield  {author} {\bibinfo {author} {\bibfnamefont {Z.}~\bibnamefont
  {Peng}}\ and\ \bibinfo {author} {\bibfnamefont {Y.}~\bibnamefont {Yang}},\
  }\href {https://doi.org/10.1360/132012-930} {\bibfield  {journal} {\bibinfo
  {journal} {SCIENTIA SINICA Physica, Mechanica $\&$ Astronomica}\ }\textbf
  {\bibinfo {volume} {43}},\ \bibinfo {pages} {579} (\bibinfo {year}
  {2013})}\BibitemShut {NoStop}%
\bibitem [{\citenamefont {St{\"u}ckelberg}(1932)}]{stuckelberg1932theory}%
  \BibitemOpen
  \bibfield  {author} {\bibinfo {author} {\bibfnamefont {E.~C.}\ \bibnamefont
  {St{\"u}ckelberg}},\ }\href@noop {} {\bibfield  {journal} {\bibinfo
  {journal} {Helv. Phys. Acta, (Basel)}\ }\textbf {\bibinfo {volume} {5}},\
  \bibinfo {pages} {369} (\bibinfo {year} {1932})}\BibitemShut {NoStop}%
\bibitem [{\citenamefont {Rabi}(1937)}]{PhysRev.51.652}%
  \BibitemOpen
  \bibfield  {author} {\bibinfo {author} {\bibfnamefont {I.~I.}\ \bibnamefont
  {Rabi}},\ }\href {https://doi.org/10.1103/PhysRev.51.652} {\bibfield
  {journal} {\bibinfo  {journal} {Phys. Rev.}\ }\textbf {\bibinfo {volume}
  {51}},\ \bibinfo {pages} {652} (\bibinfo {year} {1937})}\BibitemShut
  {NoStop}%
\bibitem [{\citenamefont {Griffiths}\ and\ \citenamefont
  {Schroeter}(2018)}]{griffiths2018introduction}%
  \BibitemOpen
  \bibfield  {author} {\bibinfo {author} {\bibfnamefont {D.~J.}\ \bibnamefont
  {Griffiths}}\ and\ \bibinfo {author} {\bibfnamefont {D.~F.}\ \bibnamefont
  {Schroeter}},\ }\href
  {https://www.cambridge.org/cn/academic/subjects/physics/quantum-physics-quantum-information-and-quantum-computation/introduction-quantum-mechanics-3rd-edition?format=HB}
  {\emph {\bibinfo {title} {Introduction to quantum mechanics, 3rd edtion}}}\
  (\bibinfo  {publisher} {Cambridge University Press},\ \bibinfo {year}
  {2018})\ pp.\ \bibinfo {pages} {403--410}\BibitemShut {NoStop}%
\bibitem [{\citenamefont {Kosugi}\ \emph {et~al.}(2005)\citenamefont {Kosugi},
  \citenamefont {Matsuo}, \citenamefont {Konno},\ and\ \citenamefont
  {Hatakenaka}}]{PhysRevB.72.172509}%
  \BibitemOpen
  \bibfield  {author} {\bibinfo {author} {\bibfnamefont {N.}~\bibnamefont
  {Kosugi}}, \bibinfo {author} {\bibfnamefont {S.}~\bibnamefont {Matsuo}},
  \bibinfo {author} {\bibfnamefont {K.}~\bibnamefont {Konno}},\ and\ \bibinfo
  {author} {\bibfnamefont {N.}~\bibnamefont {Hatakenaka}},\ }\href
  {https://doi.org/10.1103/PhysRevB.72.172509} {\bibfield  {journal} {\bibinfo
  {journal} {Phys. Rev. B}\ }\textbf {\bibinfo {volume} {72}},\ \bibinfo
  {pages} {172509} (\bibinfo {year} {2005})}\BibitemShut {NoStop}%
\bibitem [{\citenamefont {Yu}\ \emph {et~al.}(2002)\citenamefont {Yu},
  \citenamefont {Han}, \citenamefont {Chu}, \citenamefont {Chu},\ and\
  \citenamefont {Wang}}]{doi:10.1126/science.1069452}%
  \BibitemOpen
  \bibfield  {author} {\bibinfo {author} {\bibfnamefont {Y.}~\bibnamefont
  {Yu}}, \bibinfo {author} {\bibfnamefont {S.}~\bibnamefont {Han}}, \bibinfo
  {author} {\bibfnamefont {X.}~\bibnamefont {Chu}}, \bibinfo {author}
  {\bibfnamefont {S.-I.}\ \bibnamefont {Chu}},\ and\ \bibinfo {author}
  {\bibfnamefont {Z.}~\bibnamefont {Wang}},\ }\href
  {https://doi.org/10.1126/science.1069452} {\bibfield  {journal} {\bibinfo
  {journal} {Science}\ }\textbf {\bibinfo {volume} {296}},\ \bibinfo {pages}
  {889} (\bibinfo {year} {2002})}\BibitemShut {NoStop}%
\end{thebibliography}%

\end{document}